\documentclass[fleqn,usenatbib]{mnras}

\usepackage{newtxtext,newtxmath}
\usepackage{comment}

\usepackage[T1]{fontenc}
\usepackage{xcolor}

\DeclareRobustCommand{\VAN}[3]{#2}
\let\VANthebibliography\thebibliography
\def\thebibliography{\DeclareRobustCommand{\VAN}[3]{##3}\VANthebibliography}

\usepackage{graphicx}	
\usepackage{float}
\usepackage{amsmath}	
\graphicspath{{Figures/}}
\usepackage{booktabs} 
\usepackage{multirow}

\title[Abiotic Ozone Venus-like Planets]{Abiotic Ozone in the Observable Atmospheres of Venus and Venus-like Exoplanets}

\author[Robb Calder et. al.]{
Robb Calder,$^{1}$\thanks{E-mail: rdc49@cam.ac.uk}
Oliver Shorttle,$^{1,2}$
Sean Jordan,$^{3}$
Paul Rimmer,$^{2,4}$
and Tereza Constantinou$^{1}$
\\
$^{1}$Institute of Astronomy, University of Cambridge, Madingley Road, Cambridge CB3 0HA, UK\\
$^{2}$Department of Earth Sciences, University of Cambridge, Downing Street, Cambridge CB2 3EQ, UK\\
$^{3}$ETH Zurich, Institute for Particle Physics \& Astrophysics, Wolfgang-Pauli-Str. 27, 8093 Zurich, Switzerland\\
$^{4}$Cavendish Laboratory, University of Cambridge, JJ Thompson Avenue, Cambridge CB3 0HE, UK
}

\date{Accepted 2025 May 19. Received 2025 May 13; in original form 2025 March 10}

\pubyear{2025}

\begin{document}
\label{firstpage}
\pagerange{\pageref{firstpage}--\pageref{lastpage}}
\maketitle

\begin{abstract}
Ozone is a potential biosignature and disambuguator between Earth-like and Venus-like exoplanets due to its association on Earth with photosynthetically produced oxygen (O$_2$). However, the existence of ozone in Venus's observable atmosphere, a planet with no known life, raises the possibility of ozone biosignature false-positives on Venus-like exoplanets. We use a photochemical model of Venus's atmosphere to investigate the origin of its mesospheric ozone layer, and to predict how similar ozone layers would manifest for Venus-like exoplanets. For Venus, our model shows that the previously proposed fluxes of O atoms produced on the dayside and transported to the nightside cannot generate enough ozone to match the observed nightside ozone concentrations without also producing O$_2$ in excess of the observed upper limit. Nor can sufficient ozone be produced by varying the lower-atmosphere chemistry, atmospheric thermal structure, or received stellar flux in our model of Venus's atmosphere. These results imply that a presently unknown chemical pathway is responsible for the ozone production in Venus's nightside mesosphere. Ozone production rates from this pathway of 10$^5$--10$^7$ cm$^{-3}$s$^{-1}$ above the cloud layer on the nightside can re-produce the observed O$_3$ concentrations. Generalised to Venus-like exoplanets, known chemistry similarly fails to produce ozone in the abundance seen in the Venusian mesosphere. However, until the origin of Venus's ozone is understood, we cannot rule out that ozone production at concentrations observable with JWST will be common on abiotic Venus-like worlds, a possibility that limits the usefulness of ozone as a habsignature and as a biosignature.
\end{abstract}



\section{Introduction}
\label{sec:introduction}

Venus serves as a valuable case study, and potential template, for hot rocky planets: terrestrial worlds too hot to sustain liquid water, and therefore likely uninhabitable by life as we know it. Our understanding of Venus's current runaway greenhouse state \citep{Hamano2013,Boukrouche2021,Turbet2023,Constantinou2023} informs our definition of the inner edge of the habitable zone. Moreover, when transit spectroscopy rules out low mean molecular weight atmospheres for terrestrial exoplanets in the habitable zone \citep{Lustig-Yaeger2023,Lincowski2024,Rathcke2024}, Venus's CO$_2$-dominated atmosphere is a local example of a hot, high-mean molecular weight atmosphere that may exist on such planets \cite{Ducrot2024}. As a result, atmospheric models of Venus serve as a useful starting point for the atmospheres of hot rocky planets in general.

Understanding Venus's atmosphere is also crucial for observationally distinguishing between Earth-like and Venus-like exoplanets. To an extrasolar observer using transmission spectroscopy, Venus and Earth would appear outwardly similar \citep{Barstow2016,Kane2019,Jordan2021_2}. This is because of the strong CO$_2$ absorption features in the infrared (IR) that make it difficult to constrain the CO$_2$ abundance in a terrestrial exoplanet atmosphere \citep{Segura2007}, complicating the distinction between a Venusian CO$_2$-dominated atmosphere and an N$_2$-dominated atmosphere. The importance of distinguishing Venus-like planets from liquid-water habitable planets has been made clearer by \cite{Turbet2023}, who have shown that the need to condense liquid water following a planet's hot start means many liquid-water habitable zone planets may never have condensed liquid water. The \cite{Turbet2023} results suggest a very wide Venus zone that covers most of the rocky planets discovered presently. Therefore, we require atmospheric signatures that can be used to distinguish between Earth-like and Venus-like planets \citep{Barstow2016,Jordan2021}.

An example of a molecule that could be used to distinguish between Earth-like and Venus-like planets is ozone (O$_3$). In Earth's atmosphere, O$_3$ is the photochemical by-product of molecular oxygen (O$_2$), which has increased from trace amounts to 20\% of the atmosphere as a result of oxygenic photosynthesis \citep{Rutten1970}. Furthermore, unlike O$_2$, O$_3$ has a strong absorption feature in the mid-infrared range (MIR) range, specifically at 9.6$\mu$m. This wavelength range encompasses key features of potential biosignature molecules \citep{Fujii2018} and we can also measure the thermal emission of planets \citep{Greene2023} in the MIR. Detecting O$_3$ in an exoplanet atmosphere could imply the presence of O$_2$, although the relationship between their abundances is non-linear \citep{Kozaksis2022}. Therefore, O$_3$ could be used not only as a habsignature (an indicator of habitable surface conditions), but also as a potential biosignature.


However, the use of O$_3$ to distinguish between Earth-like and Venus-like exoplanets and as a biosignature is complicated by the fact that O$_3$ has been detected in Venus's atmosphere. O$_3$ has been detected above Venus's cloud layer at concentrations ranging from 0.1-1 ppm \citep{Montmessin2011, Evdokimova2021}. It has been suggested that ppm concentrations of O$_3$ could be detectable in the above-cloud atmospheres of terrestrial exoplanets with JWST \citep{Tremblay2020}. If ppm-level O$_3$ concentrations are common in the observable atmospheres of Venus-like exoplanets, then O$_3$ alone may be insufficient to distinguish between Earth-like and Venus-like exoplanets or to serve as a biosignature.


To explore the probability of observable O$_3$ in the atmospheres of Venus-like exoplanets, we must first understand the origin of the O$_3$ throughout Venus's atmosphere. \cite{Montmessin2011} reported the first direct detection of O$_3$ in Venus's mesosphere, via observations of UV absorption using the Spectroscopy for Investigation of Characteristics of the Atmosphere of Venus (SPICAV) instrument aboard Venus Express. The observations were conducted across the nightside of Venus at a range of altitudes and latitudes. The O$_3$ detections occurred on the nightside of Venus, primarily at a latitude of 30$^\circ$S, with concentrations ranging from 0.1 to 1 ppm at a mean altitude of 100\,km. \cite{Marcq2019} found evidence of another O$_3$ layer above the cloud top, also using UV absorption data from the SPICAV instrument. These observations, mainly conducted at high latitudes (above 50$^\circ$) in both hemispheres on Venus's dayside, showed concentrations ranging from 5 to 20 ppb. 

Even before the first observations of O$_3$ in Venus's atmosphere, 1D photochemical models had predicted its presence on the nightside of Venus, and suggested possible production and destruction mechanisms \citep{Sze1975,Yung1999,Krasnopolsky2010}. These models propose that O$_3$ forms from O atoms produced by CO$_2$ photolysis on Venus's dayside, which then circulate to the nightside to form O$_2$ and O$_3$. The main destruction routes for O$_3$ proposed in these studies were by chlorine (Cl), sulfur (S) and hydrogen (H) atoms. However, these models overestimated the O$_3$ concentrations observed by \cite{Montmessin2011}, therefore the model in \cite{Krasnopolsky2010} was updated to include increased fluxes of H and Cl at the upper boundary \citep{Krasnopolsky2013}. 

In response to the O$_3$ detection on the dayside by \cite{Marcq2019}, \cite{Stolzenbach2023} employed a 3D coupled climate-photochemistry model to show that this O$_3$ layer forms as a result of Hadley Cells \citep{Malkus1970}, which dominate the circulation in Venus's atmosphere at 70\,km. In the model in \cite{Stolzenbach2023}, O$_2$ is transported downwards to the cloud tops, where the higher densities facilitate the formation of O$_3$. This is in contrast to higher altitudes in Venus's atmosphere, where subsolar to anti-solar circulation dominates over circulation due to Hadley Cells \citep{Sanchez2017}. This subsolar-to-antisolar circulation is thought to generate the O$_3$ layer observed at 100,km in Venus's atmosphere, as reported by \cite{Montmessin2011}, due to the circulation of O atoms produced by CO$_2$ photolysis from the dayside to the nightside.

\begin{figure}
    \includegraphics[width=0.5\textwidth]{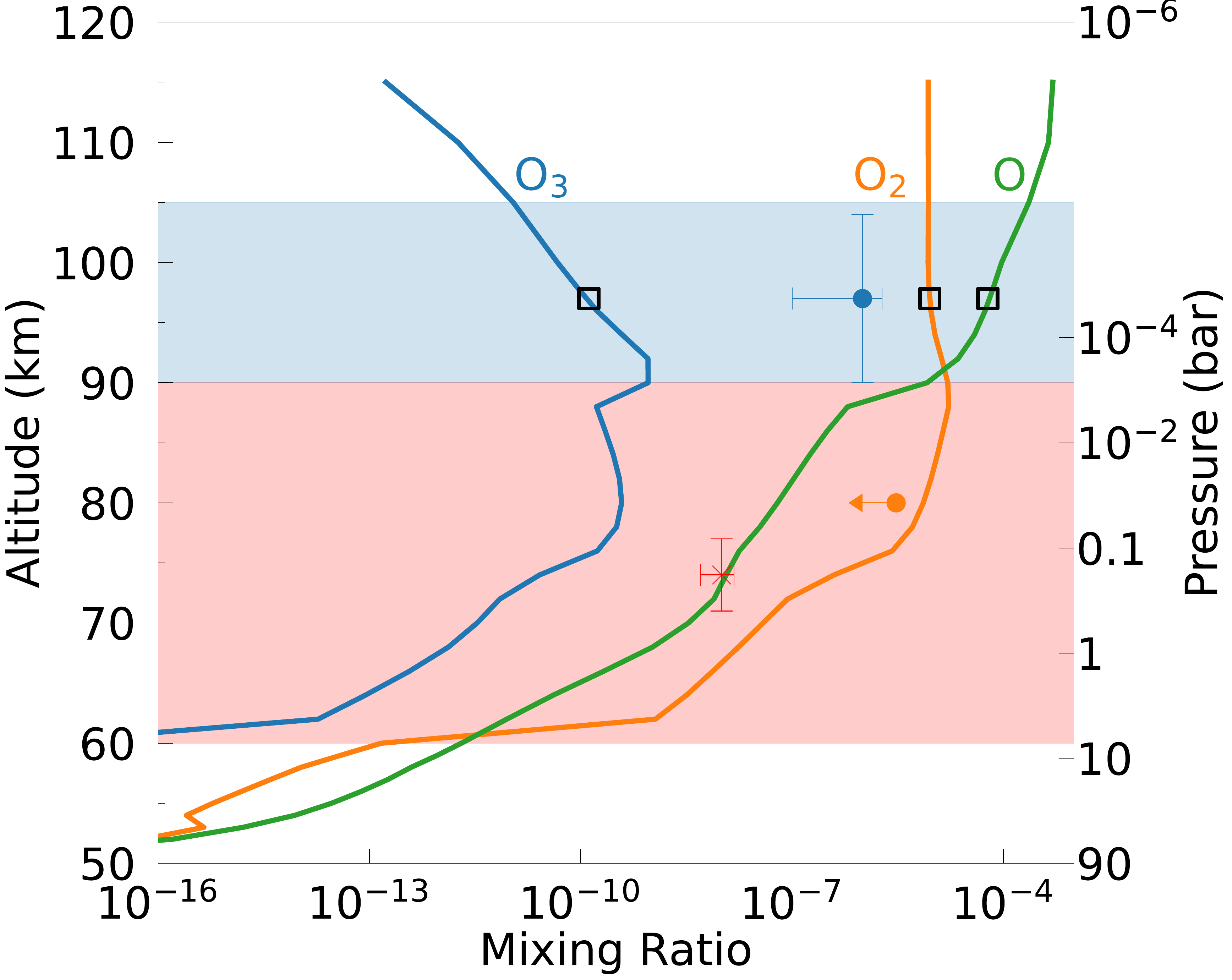}
    \caption{
        Mixing ratios for ozone ($\mathrm{O}_3$), molecular oxygen ($\mathrm{O}_2$) and atomic oxygen (O) as a function of altitude in our dayside Venus model, with the observations from \protect\cite{Montmessin2011} (blue data point) and \protect\cite{Marcq2019} (red data point), as well as the $\mathrm{O}_2$ upper limit from \protect\cite{Mills1999} (orange data point). The shaded blue and red regions corresponds to the altitude ranges for which the observations in \protect\cite{Montmessin2011} and \protect\cite{Marcq2019} were taken respectively. The black squares correspond to the mixing ratios of $\mathrm{O}_3$, $\mathrm{O}_2$ and O that are used as input to our 0D nightside chemical-kinetics model.
        }
    \label{fig:basemodels}
\end{figure}

However, the 1D photochemical studies that argue that Venus's nightside O$_3$ layer is formed from O atoms sourced from the dayside \citep{Sze1975,Yung1999,Krasnopolsky2010,Krasnopolsky2013} rely on input fluxes of radical species that are thought to circulate to the nightside from the dayside. These fluxes are based on limited observational data or inferred from column-integrated abundances in dayside models. This suggests that O atoms sourced from the dayside alone may not be able to produce the observed O$_3$ on Venus's nightside. Furthermore, dayside 1D photochemical models underestimate the nightside O$_3$ mixing ratio in the Venusian mesosphere by several orders of magnitude compared to observations, as well as overestimating the O$_2$ mixing ratio in the upper atmosphere compared to the observed upper limit reported in \cite{Mills1999} by a factor of 10 (see \cite{Mills2007,Krasnopolsky2012,Jordan2021_2} and our fiducial Venus model shown in Figure \ref{fig:basemodels}). Therefore, the origin of the nightside O$_3$ layer in Venus's mesosphere remains an unsolved problem.

The O$_3$ deficit compared to the observations reported in \cite{Montmessin2011} in photochemical models of Venus's atmosphere may be evidence of a chemical pathway that has not yet been incorporated into these models. This pathway could include salt chemistry in the clouds \citep{Rimmer2021}, cosmic ray chemistry \citep{Airey2020}, or even lightning-induced electrochemistry \citep{Qu2023}. Inaccurate cross-sections or rate coefficients in existing chemical networks may also lead to unphysically high O$_3$ or O$_2$ production or destruction rates, potentially explaining the O$_3$ deficit in current models. If the circulation of O atoms from the dayside cannot account for the O$_3$ layer in Venus's nightside, this motivates an investigation to determine if either of these possibilities can explain the O$_3$ deficit.

In this work, we use 1D photochemical models to investigate potential origins of the O$_3$ layer in the Venusian mesosphere, as well as O$_3$ production in the observable atmospheres of Venus-like exoplanets. Section \ref{sec:methods} outlines our 1D photochemical model, ARGO, and our chemical network: STAND2021, as well as the boundary conditions we use for the fiducial photochemical model of Venus's atmosphere. In section \ref{sec:chemicalcycles}, we describe the main chemical pathways involving H, S and Cl chemistry that contribute to O$_3$ formation in Venus's atmosphere. In section \ref{sec:chemicaldynamicfactors}, we describe our parameter study, in which we vary the parameters in our 1D photochemical model to explore the sensitivity of O$_3$ production in Venus's atmosphere to dynamic transport, lower-atmosphere chemistry, local temperature and stellar irradiation. In section \ref{sec:venusozonediscussion}, we assess the ability of our suite of photochemical models to explain Venus's mesospheric O$_3$ layer, and we discuss potential unknown chemical pathways that could be producing O$_3$ in Venus's mesosphere. In section \ref{sec:exoplanetscontext} we discuss the implications of our results for the use of O$_3$ as a disambiguator between Earth-like and Venus-like exoplanets as well as its use as a biosignature. We present our conclusions in section \ref{sec:conclusions}.

\section{The Photochemical Model}
\label{sec:methods}

To model the atmospheric chemistry of Venus in 1D, we use the Lagrangian photochemical-diffusion code ARGO \citep{Rimmer2016, Rimmer2021}, coupled with the ion-neutral chemical network STAND2021 \citep{Rimmer2021,Rimmer2016}. This network contains over 6000 reactions and more than 500 chemical species involving 15 elements. The rate constants in STAND2021 are valid for temperatures ranging from 100K to 30,000K \citep{Hobbs2021}, making it suitable for a variety of atmospheric contexts, from terrestrial to hot Jupiter atmospheres. The model re-produces atmospheric observations of key species in the atmospheres of Earth, Jupiter and Venus \citep{Rimmer2016,Rimmer2021}. Condensation chemistry for H$_2$SO$_4$ and other sulfur allotropes is also included.

ARGO tracks a gas parcel as it diffuses upwards and downwards through the atmosphere, solving for the parcel's chemistry at each atmospheric layer. The eddy diffusion coefficient, which describes vertical mixing, determines the timescale over which the chemistry is solved at each layer. ARGO determines the chemistry of a specific layer by solving the coupled set of non-linear differential equations that describe the number density of each chemical species as a function of time,

\begin{equation}
    \frac{dn_X}{dt}=P_X-L_Xn_X-\frac{\partial\Phi_X}{\partial z},
    \label{equ:argomainequation}
\end{equation} where $n_X$ refers to the number density of species $X$ (cm$^{-3}$). $P_X$ and L$_X$n$_X$ are the sums of the production and destruction rates of each reaction involving species $X$ respectively (cm$^{-3}$s$^{-1}$), and $\frac{\partial\Phi_X}{\partial z}$ is the divergence of the vertical diffusion flux (cm$^{-3}$s$^{-1}$), which characterises how each species mixes vertically.

The inputs to ARGO are the mixing ratio of a number of species at the base of the atmosphere, bulk planetary properties (radius, mean molecular weight etc.), the stellar spectrum incident at the top of the atmosphere, the pressure-temperature (PT) profile of the atmosphere, and the eddy diffusion profile. Once the solver has moved the parcel to the top of the atmosphere and back down to the bottom layer, it uses the stellar spectrum and the chemical abundances from the first run to calculate the incident stellar flux at each atmospheric layer. ARGO then restarts the process of moving the gas parcel upwards and downwards in the atmosphere, solving for the chemistry while accounting for photochemical reactions. The previous step is repeated until the atmospheric profiles converge. The output of the code is the mixing ratio of each species as a function of atmospheric height.

\begin{figure*}
    \includegraphics[width=\textwidth]{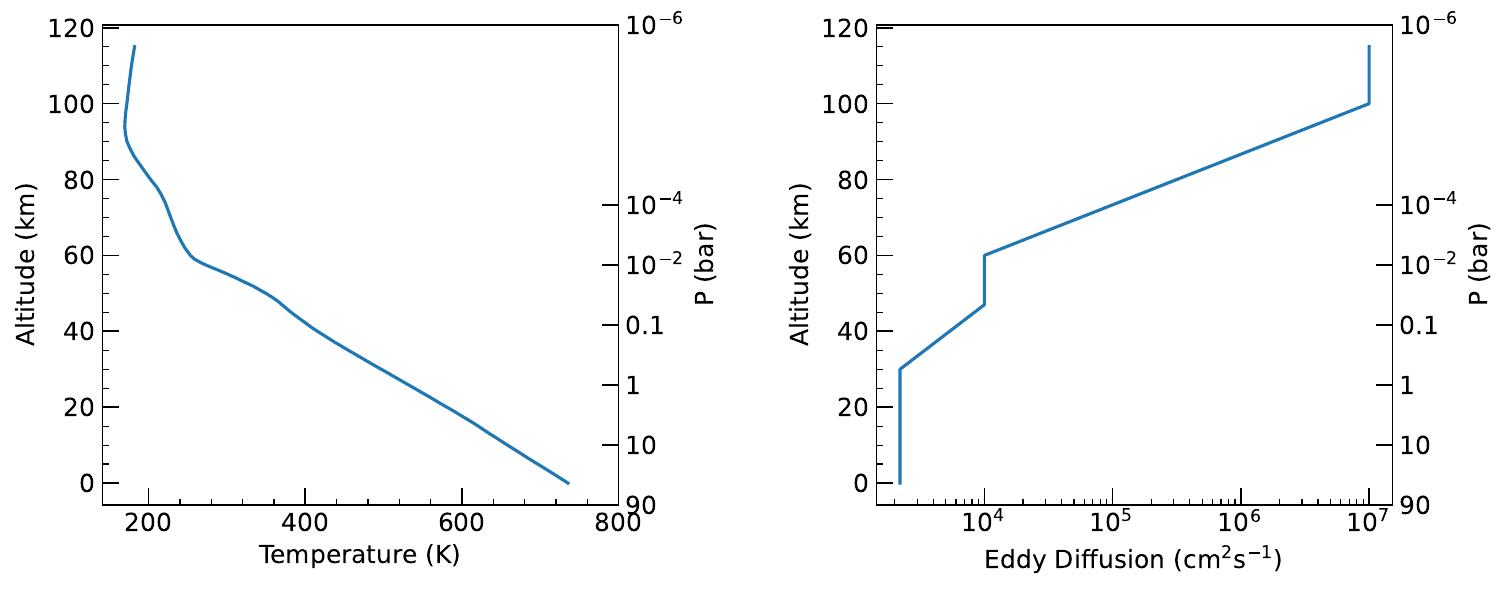}
    \caption{Pressure-temperature profile and Eddy Diffusion profile used as input for the models presented in this work. Values below altitudes of 50\,km were taken from \protect\cite{Krasnopolsky2007} and values above altitudes of 50\,km were taken from \protect\cite{Krasnopolsky2012}.}
    \label{fig:ptprofile}
\end{figure*}

In the photochemical model used in this work, we set the initial surface concentrations to 0, except for those species shown in table \ref{tab:inputconditions}. These mixing ratios are taken from \cite{Rimmer2021}. We also use three stellar spectra, one for the Sun and two for a K2.5 star (HD40307), and an M5 star (GJ581). The solar spectrum is from the SORCE data \citep{Rottman2006}, and the other stellar spectra are from the Measurements of the Ultraviolet Spectral Characteristics of Low-mass Exoplanetary Systems (MUSCLES) Treasury survey \citep{France2016}. All spectra were initially scaled to the integrated incident stellar flux (IISF) that Venus receives, using the ratio of the bolometric flux of each individual spectrum and the bolometric flux received by Venus from the Sun. Finally, we use the PT profiles and eddy diffusion profiles from \cite{Krasnopolsky2007} for altitudes between 0 and 50\,km, and the profiles from \cite{Krasnopolsky2012} for altitudes between 50\,km and 110\,km (see Figure \ref{fig:ptprofile}).

In all cases in this work, `fiducial model' refers to a photochemical model using the mixing ratios in table \ref{tab:inputconditions}, the PT profile and eddy diffusion profile shown in figure \ref{fig:ptprofile} and one of the three host stellar spectra mentioned previously. These represent planets with the same lower-atmosphere chemistry and atmospheric temperature structure as Venus around different types of host stars. Figure \ref{fig:basemodels} shows the O$_3$, O$_2$ and O mixing ratios as a function of altitude for the fiducial model around a G2 star, i.e. a simulation of Venus's dayside atmosphere. We refer to this model as the `fiducial Venus model'.

\begin{table*}
\centering
\begin{tabular}{llllllllll}
\toprule
CO$_2$ & N$_2$ & SO$_2$  & H$_2$O & CO     & OCS   & HCl     & H$_2$ & H$_2$S & NO      \\ \midrule
0.96   & 0.03  & 150 ppm & 30 ppm & 20 ppm & 5 ppm & 500 ppb & 3 ppb & 10 ppb & 5.5 ppb \\ \bottomrule
\end{tabular}
\caption{Mixing ratios of key species at the bottom of the atmosphere for our fiducial photochemical models. These values are taken from \protect\cite{Rimmer2021}.}
\label{tab:inputconditions}
\end{table*}

\section{Ozone Chemical Pathways in the Venusian Atmosphere}
\label{sec:chemicalcycles}

Throughout Venus's atmosphere, the reaction that dominates O$_3$ production is the reaction of O atoms with O$_2$,

\begin{equation}
    \mathrm{O}_2 + \mathrm{O} + \mathrm{M} \longrightarrow \mathrm{O}_3 + \mathrm{M},
    \label{chem:ozoneformation}
\end{equation} where M is a third body. Conversely, the reaction that dominates O$_3$ destruction in the upper atmosphere is its photolysis into O$_2$ and O,

\begin{equation}
    \text{O}_3 + \text{h}\nu \longrightarrow \text{O}_2 + \text{O}.
    \label{chem:ozonedissociation}
\end{equation} Therefore, the net production of O$_3$ depends on the balance between the chemical cycles producing O$_2$ and O, and the incident stellar flux. 

An important cycle producing O$_2$ and O is that involving the photolysis of CO$_2$,

\begin{equation}
    2\text{CO}_2 + 2h\nu \longrightarrow 2\text{CO}+2\text{O},
    \label{chem:CO2toCOandO}
\end{equation}
\vspace{-0.7cm}
\begin{equation}
    \text{O} + \text{O} + \text{M} \longrightarrow \text{O}_2 + \text{M},
    \label{chem:OandOtoO2}
\end{equation}
\vspace{-0.7cm}
\begin{equation*}
    \text{Net: } 2\text{CO}_2 \longrightarrow 2\text{CO}+\text{O}_2.
    \label{chem:netCO2toCOandO2}
\end{equation*}

\begin{figure*}
    \includegraphics[width=\textwidth]{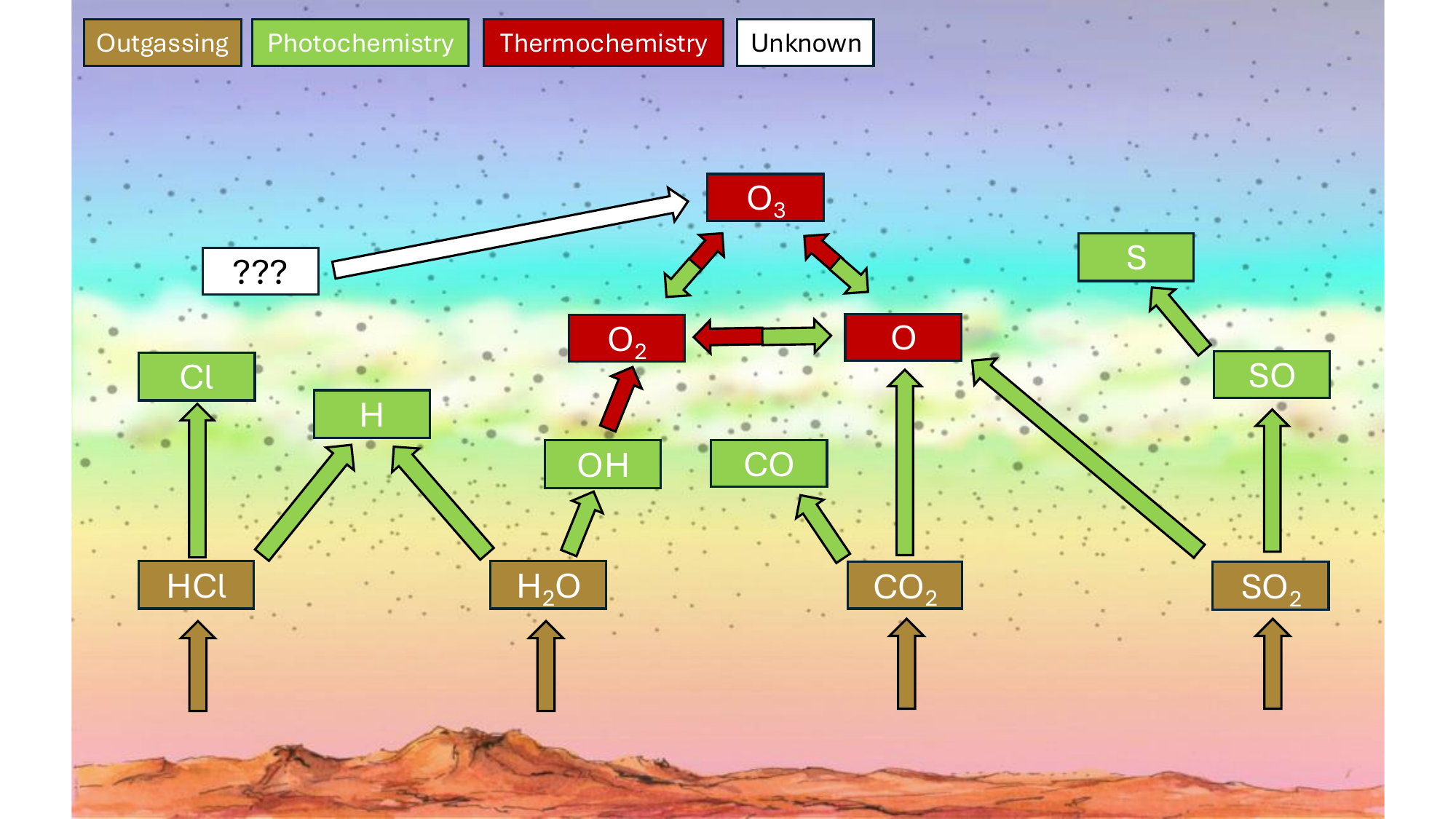}
    \caption{Schematic showing some of the major chemical cycles involved in ozone (O$_3$) formation in Venus's atmosphere. The major H, S and Cl bearing species are outgassed from Venus's surface. These species are subsequently dissociated by photochemistry, and the products are involved in chemical cycles that produce O$_2$ and O. The resulting O$_2$ and O react to produce O$_3$. We include the unknown chemical pathway producing O$_3$ required by our results.}
    \label{fig:ozoneschematic}
\end{figure*}

However, photochemical models have shown \citep{Demore1982,Yung1999,Jordan2021_2} that catalytic cycles involving Cl, H and S chemistry in Venus's atmosphere can significantly affect the production and destruction of O, O$_2$ and O$_3$ (see Figure \ref{fig:ozoneschematic} for a summary of the chemical cycles involved in O$_3$ production). We describe the main chemical cycles producing and destroying O$_3$ involving Cl-, H- and S-bearing species in turn.

Sulfur chemistry can affect O$_3$ production and destruction. The photolysis of SO$_2$ can enhance O production via,

\begin{equation}
    \text{S}\text{O}_2 + h\nu \longrightarrow \text{SO}+\text{O},
    \label{chem:SO2toSOandO}
\end{equation}
\vspace{-0.7cm}
\begin{equation}
    \text{SO} + h\nu \longrightarrow \text{S}+\text{O},
    \label{chem:SOtoSandO}
\end{equation}
\vspace{-0.7cm}
\begin{equation*}
    \text{Net: }\text{SO}_2 \longrightarrow \text{S}+2\text{O}.
    \label{chem:netSO2toSOandO}
\end{equation*} Conversely, S atoms can destroy O$_3$ and O$_2$ molecules via,

\begin{equation}
    \text{S} + \text{O}_3 \longrightarrow \text{SO}+\text{O}_2,
    \label{chem:SandO3toSOandO2}
\end{equation}
\vspace{-0.7cm}
\begin{equation}
    \text{SO} + h\nu \longrightarrow \text{S}+\text{O},
    \label{chem:SOtoSandO(2)}
\end{equation}
\vspace{-0.7cm}
\begin{equation*}
    \text{Net: }\text{O}_3 \longrightarrow \text{O}_2+\text{O},
    \label{chem:netO3toO2andO}
\end{equation*}

\begin{equation}
    \text{S} + \text{O}_2 \longrightarrow \text{SO}+\text{O},
    \label{chem:SandO2toSOandO}
\end{equation}
\vspace{-0.7cm}
\begin{equation}
    \text{SO} + h\nu \longrightarrow \text{S}+\text{O},
    \label{chem:SOtoSandO(3)}
\end{equation}
\vspace{-0.7cm}
\begin{equation*}
    \text{Net: }\text{O}_2 \longrightarrow \text{O} +\text{O}.
    \label{chem:netO2toOandO}
\end{equation*}

Hydrogen chemistry has a similar effect on O$_3$ production as sulfur chemistry. H atoms sourced from H$_2$O photolysis decrease the amount of O$_3$ in the atmosphere by reacting with O$_3$ via,

\begin{equation}
    \text{H} + \text{O}_3 \longrightarrow \text{OH}+\text{O}_2,
    \label{chem:HandO3toHOandO2}
\end{equation}
\vspace{-0.7cm}
\begin{equation}
    \text{OH} + \text{O} \longrightarrow \text{H}+\text{O}_2,
    \label{chem:HOandOtoHandO2}
\end{equation}
\vspace{-0.7cm}
\begin{equation*}
    \text{Net: }\text{O}_3 + \text{O} \longrightarrow 2\text{O}_2.
    \label{netO3andOto2O2}
\end{equation*} On the other hand, the HO$_x$ radicals enhance the production of O$_2$, and thus O$_3$, via

\begin{equation}
     \text{H} + \text{O}_2 + \text{M} \longrightarrow \text{HO}_2+\text{M},
    \label{HandO2toHO2}
\end{equation}
\vspace{-0.7cm}
\begin{equation}
     \text{HO}_2 + \text{O} \longrightarrow \text{OH}+\text{O}_2,
    \label{HO2andOtoHOandO2}
\end{equation}
\vspace{-0.7cm}
\begin{equation}
     \text{OH} + \text{O} \longrightarrow \text{H}+\text{O}_2,
    \label{HOandOtoHandO2(2)}
\end{equation}
\vspace{-0.7cm}
\begin{equation*}
     \text{Net: }\text{O} + \text{O} \longrightarrow \text{O}_2.
    \label{netOandOtoO2}
\end{equation*}

Finally, Cl atoms sourced from HCl photolysis react with O$_3$ to lower the O$_3$ abundance in the atmosphere,

\begin{equation}
    \text{Cl} + \text{O}_3 \longrightarrow \text{ClO}+\text{O}_2,
    \label{chem:ClandO3toClOandO2}
\end{equation}
\vspace{-0.7cm}
\begin{equation}
    \text{ClO} + \text{O} \longrightarrow \text{Cl}+\text{O}_2,
    \label{chem:ClOandOtoClandO2}
\end{equation}
\vspace{-0.7cm}
\begin{equation*}
    \text{Net: }\text{O}_3 + \text{O} \longrightarrow 2\text{O}_2.
    \label{chem:netO3andOto2O2(2)}
\end{equation*}

Given the numerous routes to O$_3$ production and destruction, and their dependence on S, H, and Cl chemistry, the O$_3$ concentrations in Venus's atmosphere will be sensitive to the S, H, and Cl content of the atmosphere. These chemical cycles are further influenced by factors such as dynamic transport, local temperature, and incident stellar radiation. This motivates a broad exploration of the parameter space spanned by our photochemical model to comprehensively understand O$_3$ formation in Venus's atmosphere, and thus determine the origin of the mesospheric O$_3$ layer.


\section{Chemical and Dynamical Factors Influencing Ozone Production}
\label{sec:chemicaldynamicfactors}

\subsection{Dynamical Transport of O Atoms From the Dayside}
\label{subsec:oxygencirculation}

\cite{Krasnopolsky2013} propose that the nightside O$_3$ layer in Venus's mesosphere is due to the dynamical transport of O atoms, produced from CO$_2$ photolysis on the dayside, to the nightside. In this hypothesis, the O atoms then go on to form O$_2$ and O$_3$ via reactions \ref{chem:OandOtoO2} and \ref{chem:ozoneformation} respectively. In the absence of photolysis, the O$_3$ can then potentially accumulate to higher concentrations on the nightside than on the dayside.

To explore this hypothesis, we use a combination of 1D and 0D chemical-kinetics simulations. First, we perform a 1D photochemical-kinetics simulation of the dayside atmosphere of Venus (i.e., our fiducial Venus model) to obtain a self-consistent solution for the high altitude chemistry. Then, we use the mixing ratio of each species from this model at 100\,km, the location of the \cite{Montmessin2011} observation, as input for a 0D chemical-kinetics simulation. These 0D calculations use the STAND2021 network without photochemical reactions, as well as the pressure and temperature values at 100,km from the PT profile shown in Figure \ref{fig:ptprofile}. In reality, the atmospheric temperature would be less than this value on the nightside, but in section \ref{subsec:thermochemO$_3$prod} we will show that decreasing the atmospheric temperature has a minimal effect on O$_3$ production for a planet with a G2-type host star.

In performing these simulations, we are aiming to represent a scenario where gas parcel from Venus's dayside circulating over to the nightside, evolving chemically without irradiation. We perform a series of these 0D simulations where we increase the initial O mixing ratio above that self-consistently predicted by our 1D dayside simulation. This represents increased concentrations of O at this atmospheric layer due to the proposed downward mixing of dayside O atoms \citep{Krasnopolsky2013}. In these 0D simulations, we evolve the chemistry forward in time to determine plausible amounts of O$_3$ formed on the nightside upper atmosphere on timescales comparable to Venus's atmospheric rotational timescale. Atmospheric rotational timescales were calculated using Venus's radius and wind speeds appropriate for Venus. In our simulations, we consider Venus wind speeds ranging from 10 to 1000 ms$^{-1}$ to account for order-of-magnitude outliers, given that Venusian wind speeds are of order 100 ms$^{-1}$ \citep{Sanchez2017}.



\begin{figure*}
    \includegraphics[width=0.45\textwidth]{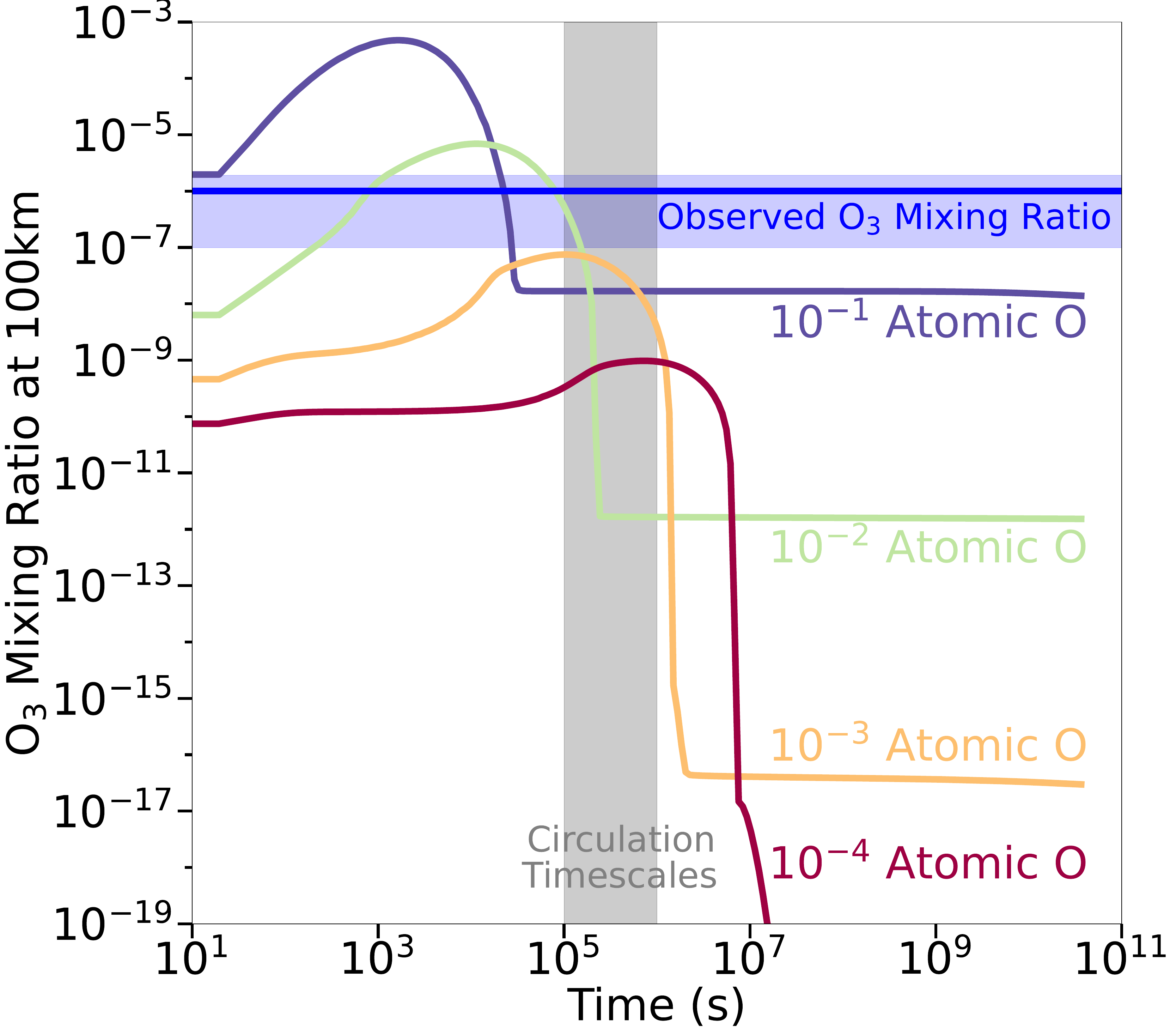}
    \includegraphics[width=0.45\textwidth]{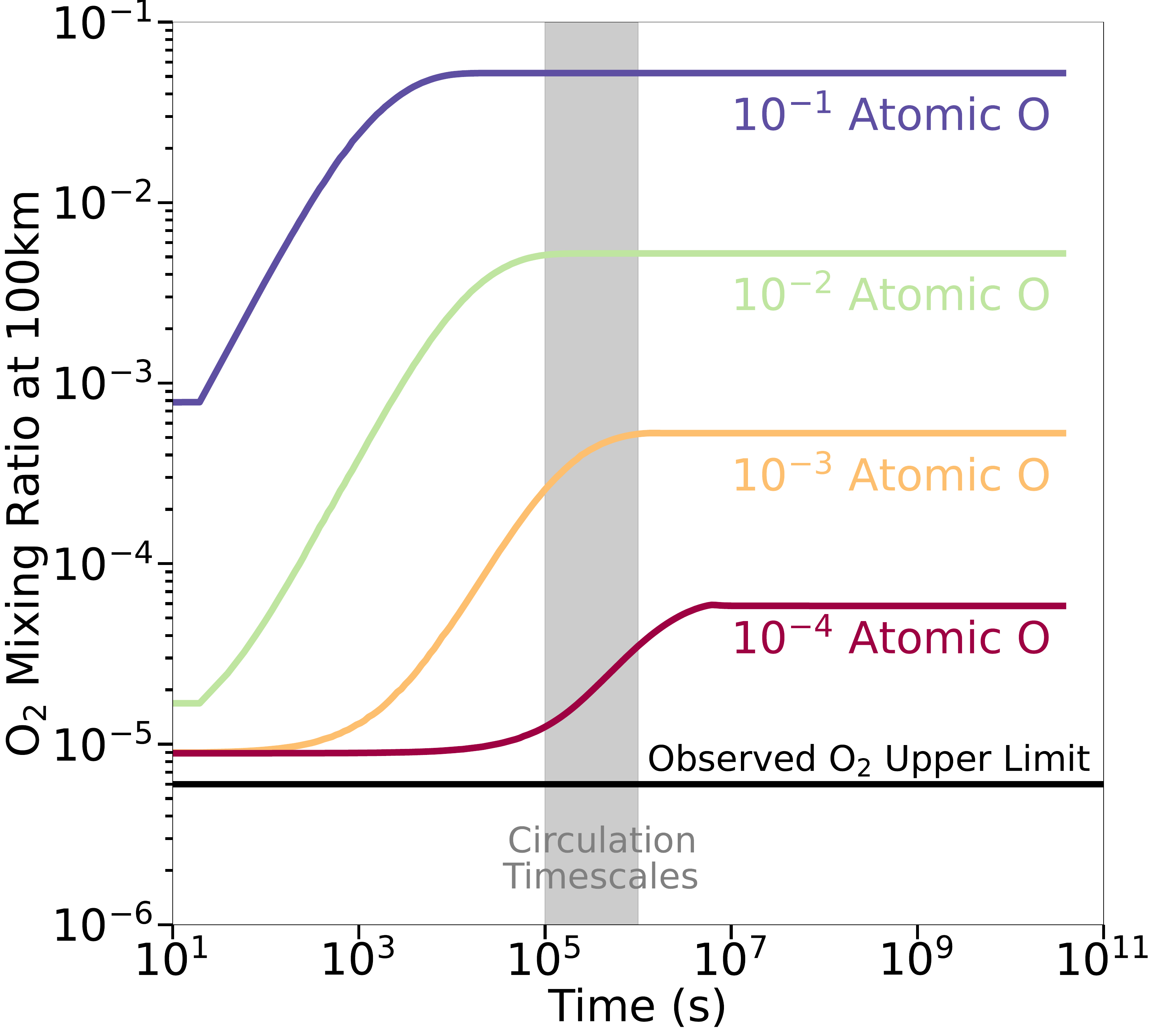}
    \caption{O$_3$ mixing ratio (left) and O$_2$ mixing ratio (right) as a function of simulation time for the 0D chemical-kinetics simulations, where the output mixing ratios at 100 km from our dayside Venus model are used as input. Photochemistry is not used in these simulations, in order to simulate the evolution of the dayside atmosphere at 100\,km as it moves across the nightside. Each line corresponds to a different O mixing ratio used as input. The blue region on the lefthand panel corresponds to the estimated O$_3$ mixing ratio with its reported uncertainty \citep{Montmessin2011} and the black line on the righthand panel corresponds to the observed upper limit O$_2$ on the O$_2$ mixing ratio. The grey region corresponds to the range of atmospheric circulation timescales for Venus's atmosphere, using minimum and maximum wind speeds of 10 ms$^{-1}$ and 1000 ms$^{-1}$ respectively.}
    \label{fig:oxygendaysidecirculation}
\end{figure*}


Figure \ref{fig:oxygendaysidecirculation} shows the results of the 0D simulations, and thus how O$_3$ concentrations and O$_2$ concentrations would vary across the nightside. O$_3$ concentrations increase initially due to the lack of photochemistry, however, once all of the O atoms are depleted by O$_2$ and O$_3$ formation, the O$_3$ is destroyed by the remaining H, S and Cl atoms. The O$_3$ concentrations then decrease until all the H, S and Cl atoms are themselves depleted by other chemical reactions, at which point the O$_3$ mixing ratio settles to a longer term stable mixing ratio \citep{Ranjan2022}. This stable mixing ratio is many orders of magnitude below the dayside O$_3$ mixing ratio, implying that the atmospheric circulation timescale on a Venus-like planet would play a large part in determining the nightside O$_3$ abundances. For all the simulations, the O$_2$ mixng ratio increases initially due to the increase in available O, but then the mixing ratio eventually saturates once there is no more O available for O$_2$ formation (reaction \ref{chem:OandOtoO2}). 

The simulations indicate that achieving ppm-level concentrations of O$_3$ at 100 km, consistent with the observations reported by \cite{Montmessin2011}, requires percent-level concentrations of atomic oxygen in the dayside atmosphere at this altitude. Additionally, they indicate that any situation in which the O mixing ratio in this layer increases leads to an even larger overestimate of the O$_2$ mixing ratio compared to the upper limit.

\subsection{Lower-Atmosphere Chemistry}
\label{subsec:surfacchemistry}

As discussed in section \ref{sec:chemicalcycles}, chemical cycles in Venus's atmosphere involving by H-, Cl- and S-bearing species influence O$_3$ production and destruction. In our photochemical model, these cycles are initiated by H-, Cl- and S-bearing minor species, whose abundances in our model are determined by the input mixing ratios shown in table \ref{tab:inputconditions}. It is possible that Venus’s lower-atmosphere chemistry could differ from that described by the mixing ratios in table \ref{tab:inputconditions}. Therefore, different abundances of H-, Cl- and S-bearing species in Venus's lower-atmosphere compared to our fiducial model are a potential explanation for the mesospheric O$_3$ layer.
 
To investigate this hypothesis, we conducted simulations in which we independently varied the mixing ratios of minor species at the bottom of the atmosphere in our photochemical model. Given the importance of S, H and Cl chemistry for O$_3$ production, as discussed in section \ref{sec:chemicalcycles}, we vary the mixing ratios of H$_2$O, SO$_2$ and HCl. These species are the most abundant H-, S- and Cl-bearing species at the base of the atmosphere in our model. These simulations were performed using the three different stellar spectra; for the solar, K2.5 and M5 stars; described in section \ref{sec:methods}. We calculate the O$_3$ column density above 70\,km, the upper-altitude of Venus’s cloud layer \citep{Titov2008}, for each model, to determine how the atmospheric O$_3$ content varies in response to changes in the parameters.

\begin{figure*}
    \includegraphics[width=\textwidth]{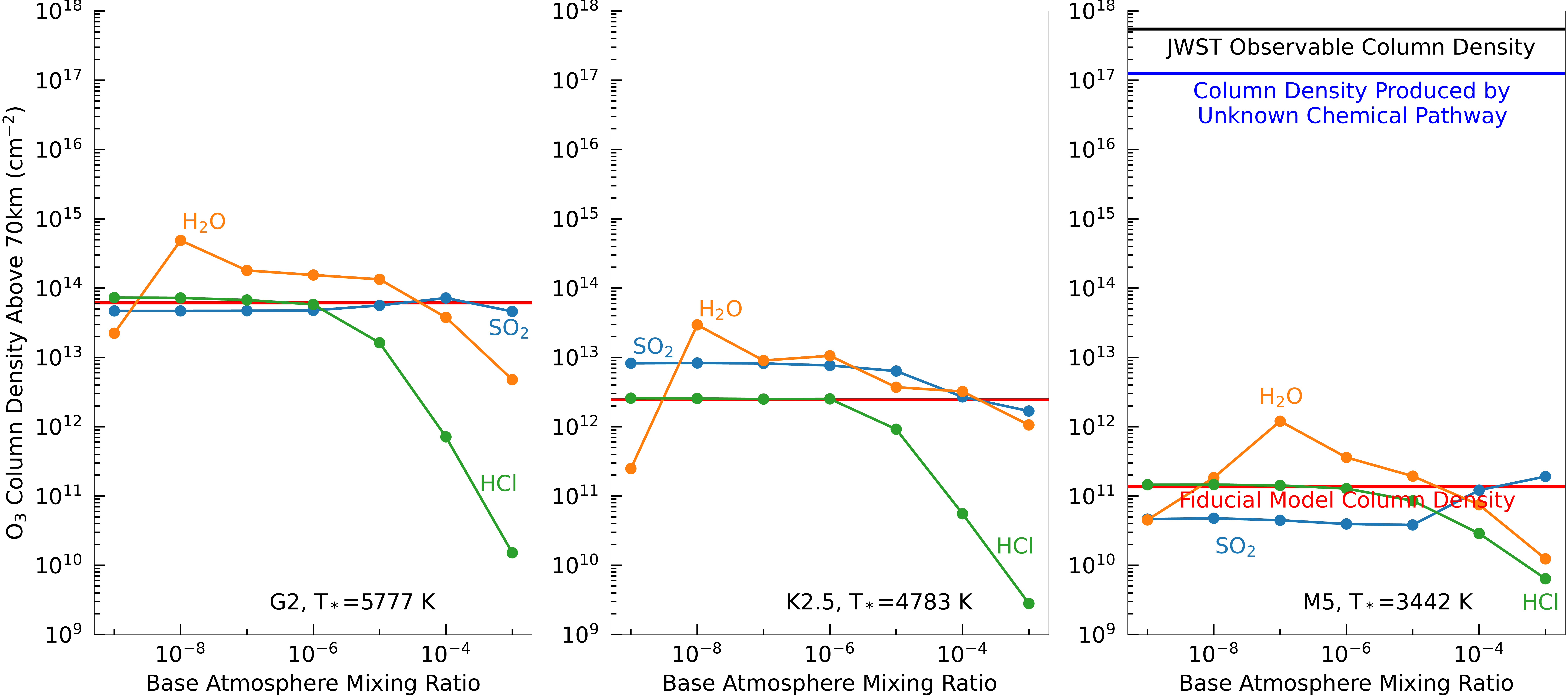}
    \caption{Ozone ($\mathrm{O}_3$) column density above 70\,km (the altitude of the top of Venus's cloud layer) as a function of the lower-atmosphere mixing ratio of minor species, each varied independently in our photochemical model. Each line represents a different minor species whose lower-atmosphere mixing ratio is varied. Each column corresponds to a different host stellar spectrum used in the model. The black line indicates the O$_3$ column density above 70\,km if the $\mathrm{O}_3$ mixing ratio were fixed at 1$\times10^{-6}$ above 70\,km, representing what would be observable in an Earth-like atmosphere with JWST according to \protect\cite{Tremblay2020}. The blue line corresponds to the $\mathrm{O}_3$ column density in our 1D photochemical model with an $\mathrm{O}_3$ flux of 10$^5$ cm$^{-3}$s$^{-1}$ injected at 64\,km, representing an unknown chemical pathway producing $\mathrm{O}_3$ in Venus's atmosphere. The red line corresponds to the relevant fiducial model: where the lower-atmosphere mixing ratios of all species are the same as those shown in table \ref{tab:inputconditions} and the appropriate stellar spectrum for the relevant column is used.}
    \label{fig:columnintegratedabundances}
\end{figure*}

Figure \ref{fig:columnintegratedabundances} shows the column density of O$_3$ above 70\,km as a function of the surface mixing ratios of H$_2$O, SO$_2$ and HCl, across different host stellar spectra. High HCl lower-atmosphere mixing ratios cause significant O$_3$ depletion throughout the atmosphere for all stellar types. For the solar spectrum, the O$_3$ column density shows little variation as the lower-atmosphere mixing ratio of SO$_2$ changes. However, at the highest SO$_2$ mixing ratios, a slight decrease in O$_3$ is observed for the K2.5 stellar spectrum, and a slight increase for the M5 stellar spectrum. Both the highest and smallest lower-atmosphere mixing ratios of H$_2$O lead to O$_3$ depletion, relative to the fiducial models, while H$_2$O mixing ratios between 10$^{-8}$ and 10$^{-5}$ result in O$_3$ enhancement.


Also, lower-mass host stars, with spectra peaking at lower wavelengths, lead to lower O$_3$ concentrations in the atmosphere, consistent with the findings of \cite{Jordan2021_2}. This trend is due to the fact that spectra for lower-mass stars are less active in the near-UV, which reduces the rate of CO$_2$ photolysis (reaction \ref{chem:CO2toCOandO}), limiting the amount of O available for O$_2$ and O$_3$ formation (reactions \ref{chem:OandOtoO2} and \ref{chem:ozoneformation}).


\subsection{Atmospheric Temperature Structure}
\label{subsec:thermochemO$_3$prod}


In Venus's upper atmosphere, the temperature structure varies significantly across different longitudes and latitudes. These variations are primarily driven by differences in solar flux received between the planet's dayside and nightside, as well as between equatorial and higher latitudes \citep{Pierrehumbert2010}. \cite{Limaye2017} show that these factors lead to atmospheric temperatures ranging from 100\,K to 300\,K in the upper atmosphere (above 78\,km) across different latitudes and local times. The chemical cycles producing O$_3$ contain thermochemical reactions (i.e. any of the reactions in section \ref{sec:chemicalcycles} that do not involve a photon), which are sensitive to local atmospheric temperature. As a result, different temperature regimes in the upper atmosphere could influence the chemical cycles responsible for O$_3$ production, and thus affect the O$_3$ concentrations in the observable atmosphere. This is another potential origin of the mesospheric O$_3$ layer. 

To explore the relationship between the atmospheric temperature structure and O$_3$ production, we vary the PT profile shown in figure \ref{fig:ptprofile} in our photochemical model. Specifically, we fix the profile to be isothermal above 78\,km, with temperatures ranging from 100K to 300K, in line with the findings of \cite{Limaye2017} (see Figure \ref{fig:differentPTprofiles} in the appendix for the different PT profiles used in this study). Although this method is a simplified approximation of the complex thermal structure in a Venus-like atmosphere, it provides valuable insight into the sensitivity of O$_3$ production pathways to local temperature. Again, these simulations were performed using the three different stellar spectra; for the solar, K2.5 and M5 stars; described in section \ref{sec:methods}.

\begin{figure}
    \includegraphics[width=0.5\textwidth]{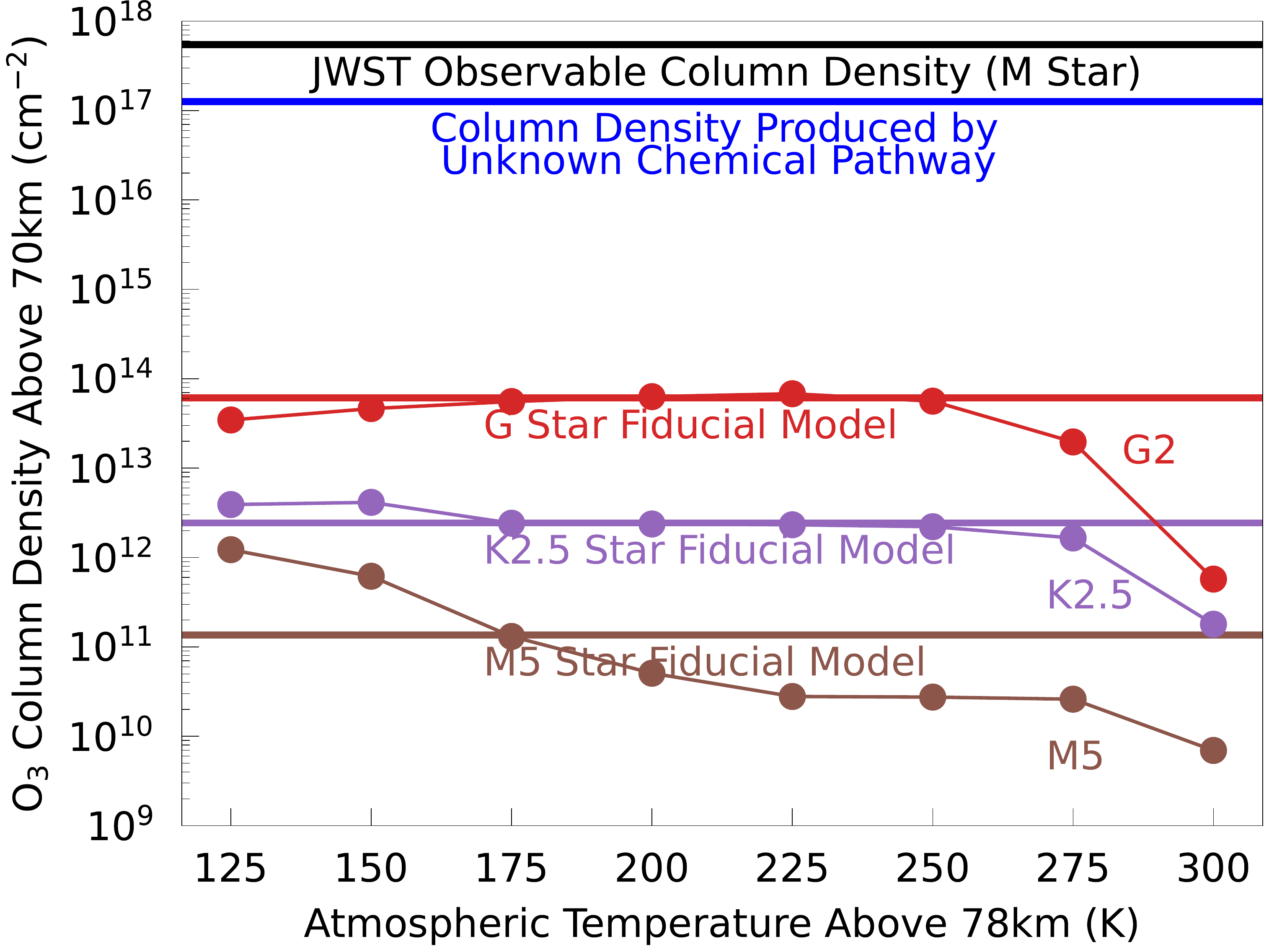}
    \caption{Ozone (O$_3$) column density above 70\,km (altitude of the Venusian cloud layer) as a function of the isothermal temperature above this altitude for the PT profiles shown in Figure \ref{fig:differentPTprofiles} in the appendix. Different line colours correspond to different host stellar specta used in the model. The black line represents the column density of O$_3$ if the O$_3$ mixing ratio were fixed at 1$\times10^{-6}$ above 70\,km. The blue line corresponds to the O$_3$ column density in our 1D photochemical model with an O$_3$ flux of 10$^5$ cm$^{-3}$s$^{-1}$ injected at 64\,km, representing an unknown chemical pathway producing O$_3$ in Venus's atmosphere. The horizontal lines indicate the O$_3$ column densities in the fiducial models: where the lower-atmosphere mixing ratios of all species are identical to those shown in table \ref{tab:inputconditions} and the pressure-temperature profile shown in figure \ref{fig:ptprofile} is used.}
    \label{fig:columnintegratedtemperature}
\end{figure}

Figure \ref{fig:columnintegratedtemperature} shows the O$_3$ column density as a function of the temperature above 78\,km for simulations using the different PT profiles. O$_3$ concentrations decrease by up to 3 orders of magnitude, compared to the fiducial model, from the coolest to the hottest temperature profile. Cooler temperatures, on the other hand, result in higher O$_3$ concentrations for the K and M star cases, but not for the solar case. This increase saturates at the lowest temperatures, however.

The inverse correlation between local atmospheric temperature and O$_3$ concentrations is due to two competing processes influencing O$_3$ production. As the temperature increases, the reaction producing O$_3$ (reaction \ref{chem:ozoneformation}) occurs at a lower rate, resulting in less O$_3$ production. As the temperature decreases, the opposite effect occurs, although the increase in O$_3$ concentrations saturates at the lowest temperatures. This saturation occurs because, as temperatures continue to drop, O$_3$ photolysis eventually dominates over thermochemistry (since photolysis reactions are independent of temperature). This limits further increases in O$_3$ concentrations. In the K- and the M-type host star cases, the reduced photolysis rate of O$_3$ allows a steeper increase in O$_3$ concentrations with decreasing temperature. However, this effect is not relevant for G-type host star case, which suggests that temperature differences between Venus's day and night side will not affect O$_3$ production significantly.


\subsection{Photochemistry}
\label{subsec:photochemo3prod}

Since O$_3$ production is strongly influenced by the production of O atoms from CO$_2$ photolysis, and that the main destruction reaction for O$_3$ in the upper atmosphere occurs through photolysis (reaction \ref{chem:ozonedissociation}), it is vital to understand how increased stellar irradiation  would affect O$_3$ concentrations in Venus's upper atmosphere.

To investigate the relationship between stellar irradiation and O$_3$ concentrations, we vary the integrated incident stellar flux (IISF) in our model by multiplying the incident stellar spectra for the solar, K2.5 and M5 stars by factors ranging from 2 to 1000. This approach was applied for each of the three stellar spectra used in this work. One caveat to this approach is that our model does not account for the additional atmospheric heating induced by the increased IISF.

\begin{figure}
    \includegraphics[width=0.5\textwidth]{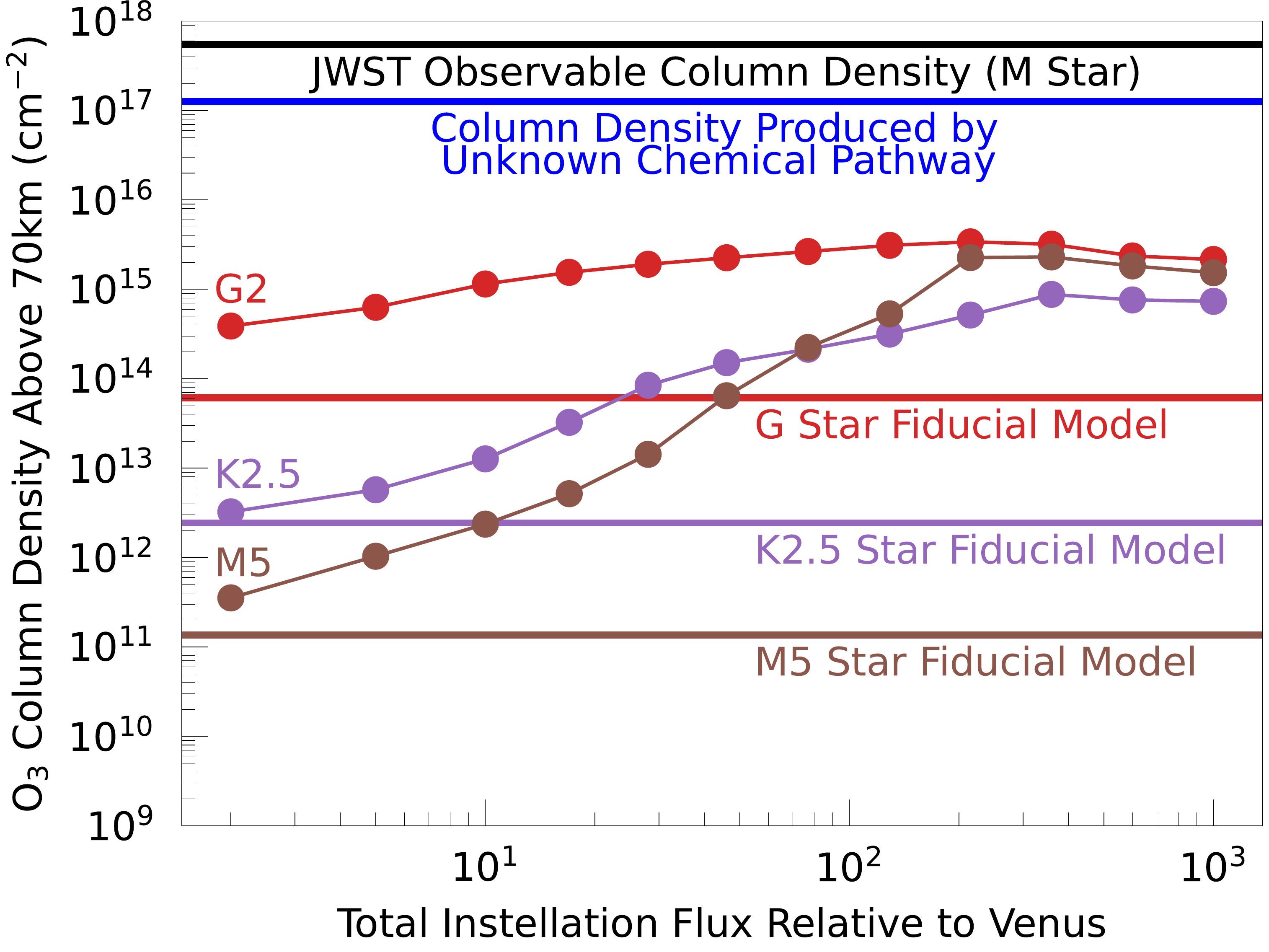}
    \caption{Column Density of ozone (O$_3$) above 70\,km (where the cloud layer is on Venus) as a function of the integrated incident stellar flux (IISF) in the simulations. Different line colours correspond to different host stellar specta used in the model. The black line represents the column density of O$_3$ if the O$_3$ mixing ratio were fixed at 1$\times10^{-6}$ above 70\,km. The blue line corresponds to the O$_3$ column density in our 1D photochemical model with an O$_3$ flux of 10$^5$ cm$^{-3}$s$^{-1}$ injected at 64\,km, representing an unknown chemical pathway producing O$_3$ in Venus's atmosphere. The horizontal lines indicate the O$_3$ column densities in the fiducial models: where the lower-atmosphere mixing ratios of all species are identical to those shown in table \ref{tab:inputconditions} and the pressure-temperature profile shown in figure \ref{fig:ptprofile} is used.}
    \label{fig:columnintegratedIISF}
\end{figure}

Figure \ref{fig:columnintegratedIISF} shows the column density of O$_3$ as a function of each IISF for the three different spectral types. A higher IISF results in increased O$_3$ concentrations for all stellar types, as a higher CO$_2$ photolysis rate leads to more O atoms available for O$_2$ and O$_3$ formation (reaction \ref{chem:ozoneformation}). This increase eventually saturates at \textasciitilde100 times Venus's IISF, due to the enhanced rate of O$_3$ photolysis (reaction \ref{chem:ozonedissociation}) with higher IISF.

The suite of photochemical models presented in this section provides a comprehensive exploration of the parameter space relevant to O$_3$ production in Venus's atmosphere. We now evaluate our models' ability to reproduce the O$_3$ concentrations reported in \cite{Montmessin2011} and, consequently, explain the origin of O$_3$ in Venus's mesosphere.

\section{The Origin of Venus's Mesospheric Ozone Layer}
\label{sec:venusozonediscussion}

\subsection{Challenges for Conventional Chemistry}

We now return to the question of the origin of the O$_3$ layer in Venus's mesosphere identified by \cite{Montmessin2011}, and utilise our photochemical models to explore various hypotheses regarding the origin of this O$_3$ layer. First, we examine the results shown in Figure \ref{fig:oxygendaysidecirculation} to determine if, as suggested in \cite{Krasnopolsky2013}, efficient O$_3$ formation can occur from from O atoms sourced from the dayside. Figure \ref{fig:basemodels} indicates that downwards transport of O atoms would need to increase the O mixing ratio at 100\,km from 10$^{-4}$ to 10$^{-2}$ to produce enough O$_3$ to match the nightside concentrations reported by \cite{Montmessin2011}. The model also suggests that, if this were achieved, the resulting O$_3$ is likely to survive, in the absence of photochemistry, over timescales consistent with Venus's atmospheric circulation timescale.

The key question, then, is whether atmospheric mixing processes can drive sufficient downward transport of O atoms to 100\,km, increasing the O mixing ratio from 10$^{-4}$ to 10$^{-2}$. Diffusive mixing, whereby O atoms would diffuse from areas of high concentration to areas of low concentration, could achieve this if the O mixing ratio at higher altitudes were to reach percent concentrations. However, our model shows that at altitudes above 100\,km, the O mixing ratio never exceeds 10$^{-3}$ (Figure \ref{fig:basemodels}). This is in agreement with the O mixing ratios predicted at high latitudes in 3D models \citep{Gilli2017} as well as O mixing ratios inferred from dayglow mapping of Venus's dayside \citep{Soret2012_b}. Furthermore, the O column-density in our dayside model is in agreement to within an order of magnitude with that observed on the dayside of Venus \citep{Hubers2023}. Even if the O mixing ratio were to reach percent concentrations at altitudes above the highest altitude in our model, 130\,km, the increase in O mixing ratio would be offset by the corresponding decrease in O number density at such high altitudes. This would result in very little net increase of O atoms available for O$_2$ and O$_3$ formation (reactions \ref{chem:OandOtoO2} and \ref{chem:ozoneformation} respectively).

Another atmospheric mixing process that could increase the O mixing ratio at 100\,km from 10$^{-4}$ to 10$^{-2}$ is the operation of Hadley cells in Venus's atmosphere. Hadley cells occur where warm air in the equator rises to higher latitudes, then descends and flows back to lower altitudes \citep{Malkus1970}. They are thought to form the O$_3$ layer observed by \cite{Marcq2019} at 70\,km by bringing O$_2$ from higher altitudes down to 70\,km, where O$_3$ then forms via reaction \ref{chem:ozoneformation} \citep{Stolzenbach2023}. However, \cite{Stolzenbach2023} do not expect Hadley cells to operate in the Venusian mesosphere, and therefore, Hadley Cells are unlikely to account for the mesospheric O$_3$ layer. At these altitudes, sub-solar to anti-solar circulation is expected to dominate \citep{Sanchez2017}.

Even if an atmospheric mixing process could increase the O mixing ratio at 100 km from 10$^{-4}$ to 10$^{-2}$, without a simultaneous mechanism to remove O$_2$, this would further elevate the O$_2$ mixing ratio at this altitude beyond the observed upper limit (figure \ref{fig:oxygendaysidecirculation}). The O$_2$ mixing ratio predicted by our model at this altitude is in agreement with that predicted by other photochemical models \citep{Mills2007,Krasnopolsky2012}. For these reasons, we argue that the hypothesis that the O$_3$ layer on Venus's nightside forms purely from O atoms transported from the dayside is unlikely.


\cite{Krasnopolsky2013} argue that the O$_3$ layer on Venus's nightside can be formed at the required mixing ratio due to chemistry initiated by radical species that originate from the dayside, contrary to our findings. In their model, O$_3$ formation due to O atoms sourced from the dayside (reaction \ref{chem:ozoneformation}) and O$_3$ destruction due to H and Cl atoms from the nightside (reactions \ref{chem:ClandO3toClOandO2} and \ref{chem:HandO3toHOandO2} respectively), act to produce the O$_3$ mixing ratio observed in \cite{Montmessin2011}. They justify the values for the fluxes of these radical species used as input to their model on the basis that they are proportional to the column densities of these species in the daytime model from \cite{Krasnopolsky2012}. However, \cite{Krasnopolsky2012} do not explicitly report the values of these column densities, nor do they detail the method by which the fluxes are inferred from the column rates, making it challenging to directly compare their model with ours. Also, the model in \cite{Krasnopolsky2013} only accounts for fluxes of O, CO, H and Cl from the dayside to the nightside; a self-consistent treatment would account for the fluxes of all radicals produced on the dayside of Venus's. These fluxes would be calculated using the results of a dayside 1D photochemical model of Venus's atmosphere.

\begin{figure}
    \includegraphics[width=0.5\textwidth]{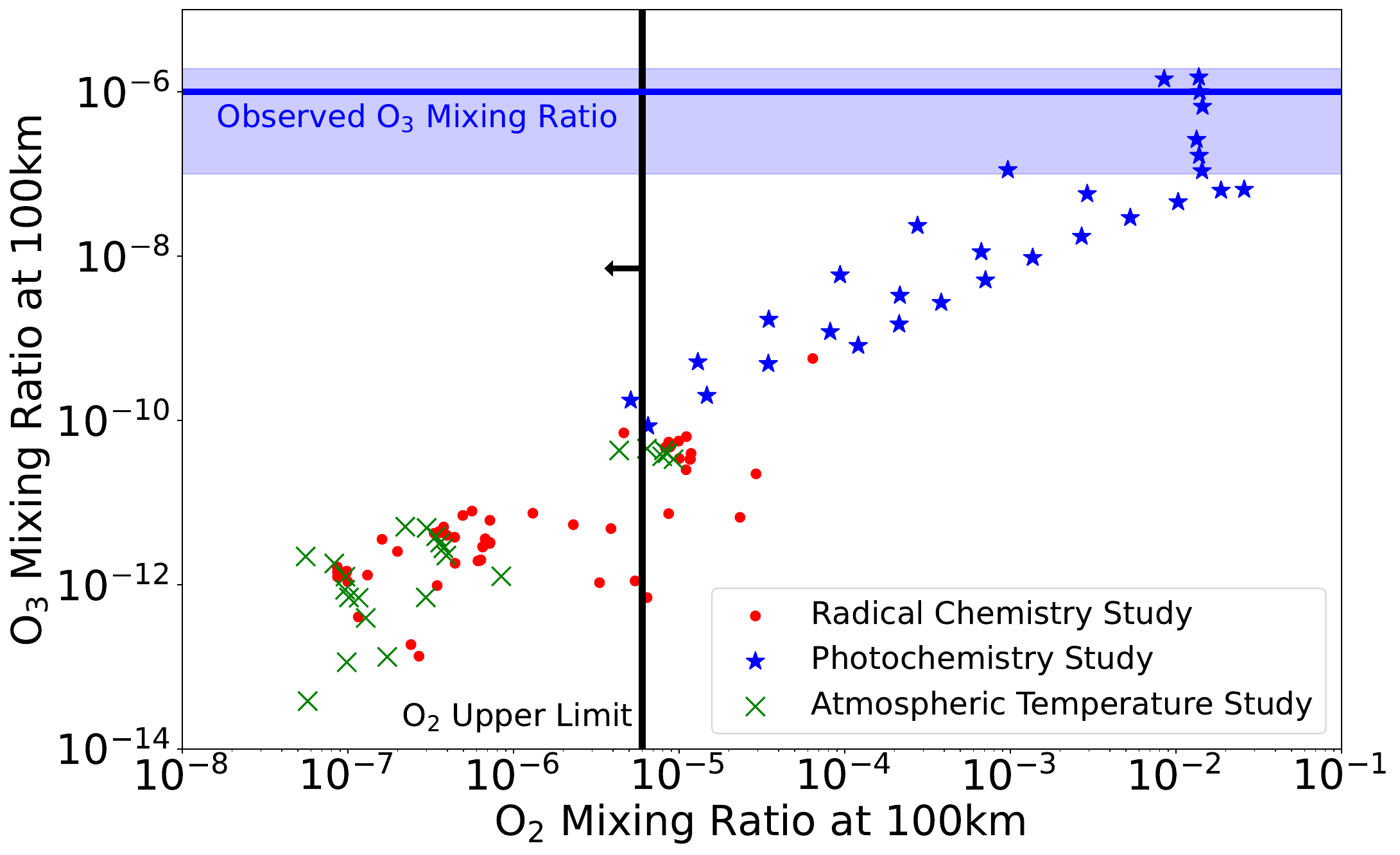}
    \caption{The $\mathrm{O}_3$ mixing ratio as a function of the $\mathrm{O}_2$ mixing ratio at 100\,km for all of the 1D simulations of Venus's atmosphere. The colour of each point corresponds to the study to which the models belong. The shaded region indicates the uncertainty in the O$_3$ mixing ratio observed in \protect\cite{Montmessin2011}. The horizontal lines represent the observational upper limit on the $\mathrm{O}_2$ mixing ratio from \protect\cite{Mills1999}.}
    \label{fig:o3vso2}
\end{figure}

With our model disfavoring the hypothesis that nightside O$_3$ forms from O atoms transported from the dayside, we now turn to the 1D photochemical models that span the parameter space relevant for O$_3$ production in Venus's atmosphere. We only consider the models that use a solar spectrum for our discussion of Venus. These models have varying abundances of minor species at the base of the atmosphere, atmospheric temperature structures and stellar radiation fluxes. For any of these models to be able to explain Venus's mesospheric O$_3$ layer, they must achieve a 1ppm O$_3$ mixing ratio at 100\,km and an O$_2$ mixing ratio less than 6ppm at the same altitude. However, Figure \ref{fig:o3vso2} shows that neither the models in which the lower-atmosphere chemistry is varied, nor the models in which the atmospheric temperature structure is varied, can reproduce the observed O$_3$ concentrations. The only models that can re-produce the observed O$_3$ concentrations do so at IISFs inconsistent with modern Venus by a factor of 100. Therefore, none of variations in lower-atmosphere chemistry, changes in atmospheric temperature structures, or increased stellar flux can explain the O$_3$ layer in Venus's mesosphere.


The primary issue with these models is their inability to simultaneously satisfy the observational constraints for O$_3$ and O$_2$. Figure \ref{fig:o3vso2} shows a clear relationship between the two mixing ratios, evidencing the difficulty of raising the O$_3$ mixing ratio to the required concentration without also increasing the O$_2$ mixing ratio beyond the observed upper limit. This relationship is unsurprising, given that the main production pathway for O$_3$ requires O$_2$ (\ref{chem:ozoneformation}) and that the main destruction pathway for O$_3$ (\ref{chem:ozonedissociation}) also produced O$_2$. 

These results leave several possible explanations for inability of our photochemical model to match the observational constraints for O$_3$ and O$_2$. It is possible that the observational constraints are themselves inaccurate. Also, the rate constants or cross-sections pertaining to key reactions in our chemical network may be inaccurate. However, another possibility is that there is an as-of-yet unconsidered chemical pathway operating in Venus's atmosphere. Such a pathway would need to produce the correct balance of O$_3$ and O$_2$, through a combination of O$_3$ production and O$_2$ destruction, allowing both observational constraints to be satisfied.

\subsection{Ozone Production Rates Required by an Unknown Chemical Pathway}
\label{subsec:requiredcloudproductionrate}

We have shown that the existing chemical reactions in our network cannot simultaneously match the O$_3$ and O$_2$ mixing ratios, even allowing for wide variation in the physical and chemical properties of the atmosphere. We now use our 1D photochemical model to estimate the O$_3$ production flux that would raise the O$_3$ concentration in the upper atmosphere to the concentration required by observations. We also determine the optimum altitude at which to produce this O$_3$, and compare the lifetime of the resulting O$_3$ on the dayside compared to the nightside.

\begin{figure*}
    \includegraphics[width=\textwidth]{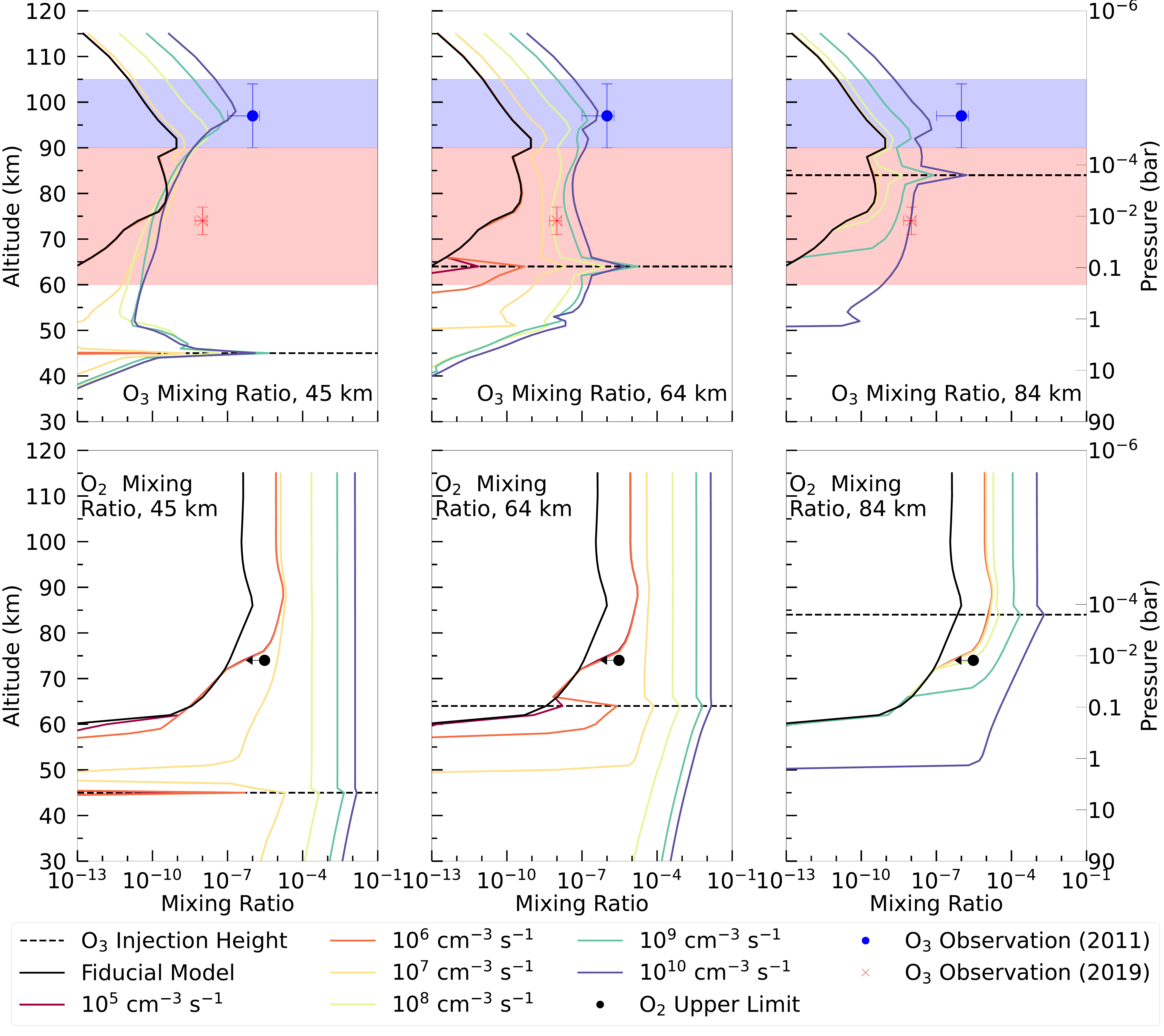}
    \caption{Dayside ozone ($\mathrm{O}_3$) mixing ratios (top row) and molecular oxygen ($\mathrm{O}_2$) mixing ratios (bottom row) as a function of altitude for fiducial Venus models with an arbitrary flux of $\mathrm{O}_3$ introduced into the dayside atmosphere. Each line represents a different $\mathrm{O}_3$ flux into the atmosphere (in cm$^{-3}$s$^{-1}$). The columns indicate the various altitudes at which the $\mathrm{O}_3$ flux is applied. Data points from observations by \protect\cite{Montmessin2011} (blue) and \protect\cite{Marcq2019} (red) are also shown, along with the altitude ranges corresponding to each observation (blue and red, respectively). The $\mathrm{O}_2$ upper limit from \protect\cite{Mills1999} is also shown.}
    \label{fig:ghosto3photo}
\end{figure*}

\begin{figure*}
    \includegraphics[width=\textwidth]{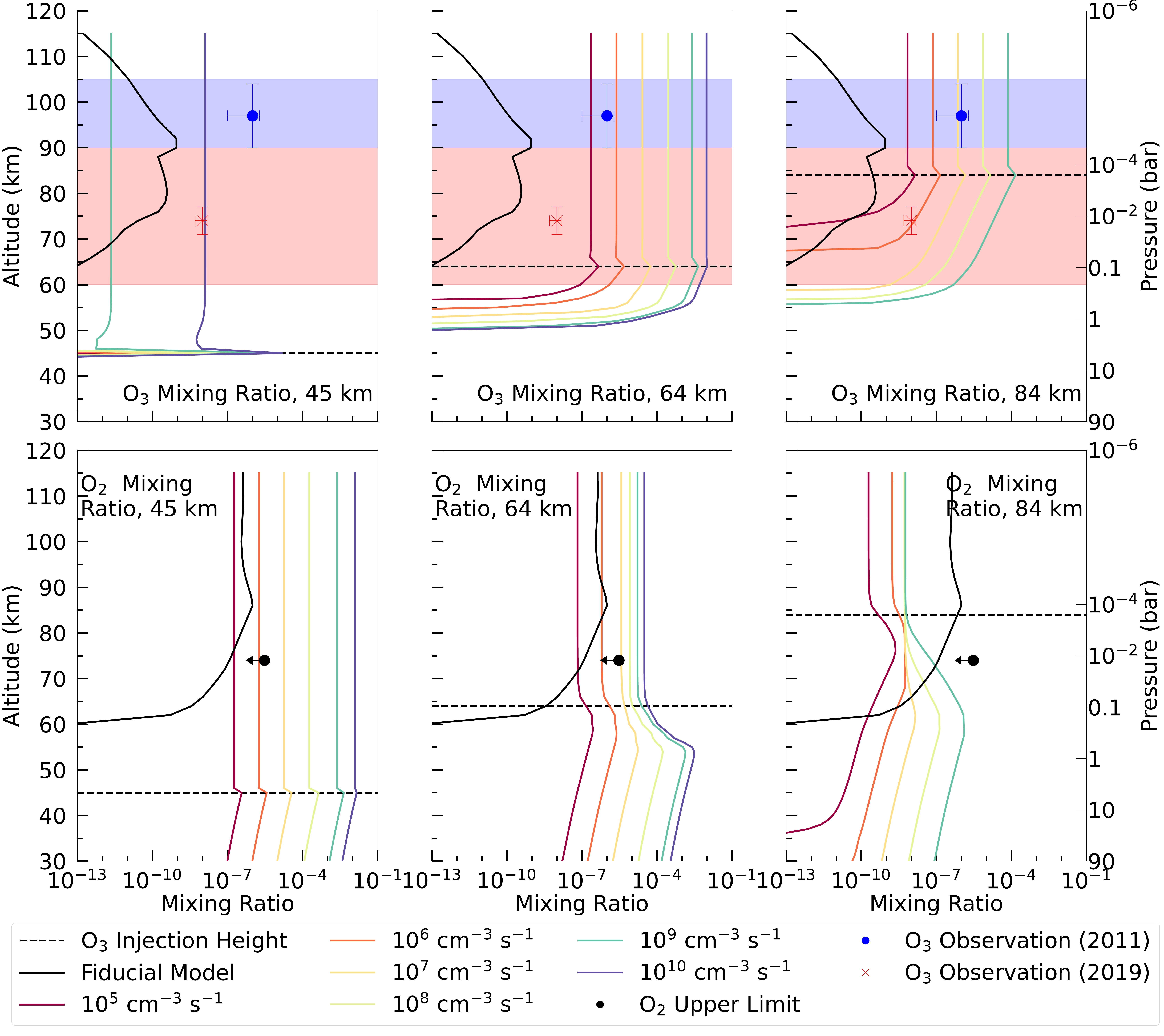}
    \caption{Nightside ozone ($\mathrm{O}_3$) mixing ratios (top row) and molecular oxygen ($\mathrm{O}_2$) mixing ratios (bottom row) as a function of altitude for fiducial Venus models with an arbitrary flux of O$_3$ introduced into the dayside atmosphere. Each line represents a different $\mathrm{O}_3$ flux into the atmosphere (in cm$^{-3}$s$^{-1}$). The columns indicate the various altitudes at which the $\mathrm{O}_3$ flux is applied. Data points from observations by \protect\cite{Montmessin2011} (blue) and \protect\cite{Marcq2019} (red) are also shown, along with the altitude ranges corresponding to each observation (blue and red, respectively). The $\mathrm{O}_2$ upper limit from \protect\cite{Mills1999} is also shown.}
    \label{fig:ghosto3nophoto}
\end{figure*}

To determine the O$_3$ production rate necessary to raise the O$_3$ concentration in the upper atmosphere to that observed by \cite{Montmessin2011}, we introduce an arbitrary flux of O$_3$ at a specific altitude in our fiducial Venus model. We conducted simulations both with and without photochemical reactions to model O$_3$ production via this pathway on the dayside and the nightside of Venus respectively. We use the same PT profile for the dayside and nightside simulations, given that decreasing the atmospheric temperature has minimal effect on O$_3$ concentrations for a planet with a G-type host star (figure \ref{fig:columnintegratedtemperature}). Figures \ref{fig:ghosto3photo} and \ref{fig:ghosto3nophoto} show the O$_3$ and O$_2$ mixing ratios as functions of altitude for a range of arbitrary O$_3$ fluxes at different altitudes, corresponding to the dayside and nightside simulations respectively. 

In the dayside models, adding O$_3$ at altitudes of 45\,km and 65\,km can significantly alter the profiles of O$_3$ and O$_2$. As shown in Figure \ref{fig:ghosto3photo}, for fluxes of 10$^5$ and 10$^6$ cm$^{-3}$s$^{-1}$, most of the additional O$_3$ is destroyed before it can be transported by vertical mixing. However, for higher fluxes, sufficient O$_3$ is added that it survives over timescales longer than the vertical mixing timescale, resulting in enhanced O$_3$ concentrations in the lower-atmosphere. In contrast, when O$_3$ is added at 85\,km, while O$_3$ concentrations in the upper atmosphere increase, none of it reaches the lower-atmosphere. For fluxes of 10$^9$ and 10$^{10}$ cm$^{-3}$s$^{-1}$ introduced at altitudes of 45\,km and 60\,km, the O$_3$ concentrations in the upper atmosphere increase sufficiently to match the SPICAV observation at 100\,km. However, these O$_3$ fluxes also raise the O$_2$ mixing ratio to 10$^{-3}$, which violates the upper limit for O$_2$. The O$_3$ concentrations in these models also exceed the observations reported by \cite{Marcq2019} at 75\,km. Therefore, O$_3$ production on the dayside cannot achieve the correct balance between O$_3$ and O$_2$.

Alternatively, when considering the models of O$_3$ production on the nightside (Figure \ref{fig:ghosto3nophoto}), the observational constraints are much easier to satisfy. Due to the chemical lifetime of O$_3$ and O$_2$ being much longer in the absence of photolysis, the additional O$_3$, as well as the O$_2$ that is formed from the thermochemical decay of the additional O$_3$, is immediately quenched upon its addition to the atmosphere in all cases. For O$_3$ production at 65\,km (within the cloud layer), production rates ranging from 10$^5$ to 10$^6$ cm$^{-3}$s$^{-1}$ can achieve the desired O$_3$ mixing ratio. If O$_3$ is produced at 85\,km (above the cloud layer), a higher production rate of 10$^7$ cm$^{-3}$s$^{-1}$ is required. Fundamentally, however, the O$_2$ mixing ratios that result from this range of O$_3$ production rates do not violate the observed upper limit. This finding suggests that, if O$_3$ is being produced in or above the nightside Venusian cloud layer with a flux of 10$^5$--10$^7$ cm$^{-3}$s$^{-1}$, then it can accumulate to concentrations consistent with those observed by \cite{Montmessin2011} without violating the observational constraint on O$_2$.

A caveat, however, in our nightside model is that it does not account for the flux of radicals from the dayside onto the nightside. Nevertheless, this omission would not substantially affect our results. In section \ref{subsec:oxygencirculation}, we demonstrated that, while destruction of O$_3$ on the nightisde due to H and S atoms is significiant, it occurs over timescales equal to or longer than atmospheric circulation timescales, therefore it is unlikely that H and S atoms could destroy O$_3$ immediately if it is being continuously replenished from a source in the cloud layer. Additionally, Figure \ref{fig:ghosto3nophoto} shows that the O$_2$ mixing ratio that results from O$_3$ fluxes of 10$^5$--10$^6$ cm$^{-3}$s$^{-1}$ is comparable to or larger than the O$_2$ mixing ratio on the dayside. This suggests that the sum of the flux of O$_2$ from the dayside and that produced on the nightside is unlikely to result in an O$_2$ mixing ratio greater than the observed upper limit, nor is the flux of O$_2$ from the dayside likely to lead to significantly increased O$_3$ formation above what we are proposing.

Another caveat is that we consider only the production of O$_3$ by an unknown chemical pathway. It is possible that this pathway could act as a sink of O$_2$ as well as a source of O$_3$, or that another pathway altogether could act as a sink of O$_2$. This would then allow higher rates of O$_3$ production by an unknown pathway without violating the observational constraint on the O$_2$ mixing ratio. Furthermore, a sink of O$_2$ could allow buildup of O$_3$ on Venus's nightside to the levels observed in \cite{Montmessin2011} due to O atoms sourced from the dayside, as discussed in \ref{subsec:oxygencirculation}, without violating the observational constraint on O$_2$ from \cite{Mills1999}. However, a mechanism for achieving percent-level O$_3$ concentrations on Venus's nightside would still be required. To the best of our knowledge, all known O$_2$ destruction pathways are accounted for in our model, therefore an as-of-yet undiscovered chemical pathway is required.

\subsection{Lightning Produced O$_3$}
\label{subsec:lightningozone}


If an additional source of O$_3$ is required, that is not captured by the reactions in our present chemical network, then one possibility is lightning-induced electrochemistry, as suggested by the experimental results of \cite{Qu2023}. Although direct observational evidence of lightning and related electromagnetic emissions in Venus's atmosphere has produced contradictory conclusions \citep{Krasnopolsky1980, Gurnett2001, Lorenz2019}, indirect evidence for lightning comes from the detection of NO in the lower atmosphere at concentrations of 5.5$\pm$1.5 ppb \cite{Krasnopolsky2006_2}. This is noteworthy because NO is challenging to produce through thermochemistry or photochemistry below the cloud layer. In this scenario, N$_2$ is dissociated via electrochemistry during lightning. The N atoms subsequently react with CO$_2$ or O atoms to produce NO \citep{Delitsky2015,Qu2023}. Therefore, lightning is thought to be the only source of NO below the clouds.


Gas-discharge experiments conducted by \cite{Levine1982} yielded 3.7 $\pm$ 0.7 $\times$ 10$^{15}$ and 4 $\times$ 10$^{17}$ molecules of NO and CO per joule of lightning, respectively. Using this yield, \cite{Krasnopolsky2006_2} deduced that a lightning rate of 0.19 $\pm$ 0.06 erg cm$^{-2}$ s$^{-1}$ is necessary to maintain the observed NO concentrations in the atmosphere, counteracting atmospheric sinks. While there are no experimentally-measured yields of O$_3$ from lightning under Venus-like conditions, CO yield can serve as a proxy for the O$_3$ yield. \cite{Qu2023} determined O$_3$ and CO concentrations in Venusian lightning discharge experiments, which can be combined with the \cite{Levine1982} CO yield. If the mixing ratios of O$_3$ and CO from the experiment are assumed proportional to the flux from lightning, the O$_3$ flux from lightning can be estimated at 5.5 $\times$ 10$^{6}$ cm$^{-2}$ s$^{-1}$.

If we consider our nightside simulation, where O$_3$ is injected in the atmosphere at 65\,km at a rate of 10$^5$ cm$^{-3}$s$^{-1}$, we can numerically integrate this O$_3$ production rate throughout the atmosphere to obtain the column O$_3$ production rate required by our simulations. This calculation yields a value of 4.4 $\times$ 10$^{10}$ cm$^{-2}$ s$^{-1}$. This value is 4 orders of magnitude greater than what is expected from lightning, suggesting that the O$_3$ production rate due to lightning is insufficient to meet the required O$_3$ concentrations. For this reason, lightning is unlikely to be a major source of O$_3$ in Venus's mesosphere.

With the question of O$_3$ in Venus's observable atmosphere explored to the extent possible with our models, we are now better positioned to address the question of O$_3$ in the atmospheres of Venus-like exoplanets. Although we have not fully answered the question of the unknown chemical pathway responsible for Venus's mesospheric O$_3$ layer, the O$_3$ production rates required from this pathway offer valuable context for evaluating other potential O$_3$ production mechanisms in Venus-like exoplanet atmospheres.

\section{Ozone in the Observable Atmospheres of Venus-like Exoplanets}
\label{sec:exoplanetscontext}

\subsection{The Relationship Between Ozone Concentrations and Conventional Chemistry}
\label{subsec:ozoneconventionalchemistryexoplanets}

We now turn to the question of using O$_3$ as a disambiguator between Earth-like and Venus-like exoplanets, as well as the potential for an O$_3$ biosignature false-positive in the atmosphere of a Venus-like exoplanet. If an Earth-like exoplanet around an M star were to have above-cloud O$_3$ at ppm concentrations, then this could be detectable with approximately 50 transits of JWST \citep{Tremblay2020}. Given the expected similarity in transmission spectroscopy between Earth-like and Venus-like exoplanets, it is reasonable to expect that ppm O$_3$ concentrations in a Venus-like exoplanet's upper atmosphere could be detectable with a similar number of transits with JWST, thus complicating the use of O$_3$ as a disambiguator. Therefore, we require an understanding of the expected O$_3$ content of Venus-like exoplanet atmospheres, as well as the probability of these exoplanets having potentially observable O$_3$ in their upper atmospheres.


We use the suite of 1D photochemical models described in section \ref{sec:chemicaldynamicfactors} to explore the parameter space of Venus-like exoplanets relevant to O$_3$ production. Our goal is to identify regions within this parameter space where O$_3$ concentrations could reach potentially observable, i.e. ppm, levels. Specifically, we analyze the results from simulations that vary: (1) the abundances of key species in the lower-atmosphere, (2) the pressure-temperature structure of the atmosphere, and (3) the IISF.

\textit{Lower-Atmosphere Chemistry}: The abundances of minor H-, Cl- and S-bearing species in the lower-atmospheres of Venus-like exoplanets will vary considerably, due to the different planet formation environments \cite{Mandell2007,Bond2010,Adibekyan2021} and outgassing mechanisms \cite{Liggins2022,Liggins2023,Guimond2023} that would exist for these planets. It is therefore important to consider how this variability in the lower-atmosphere influences the O$_3$ content in the observable upper atmospheres of Venus-like exoplanets. Our results show that, within the paradigm of our modeling framework, Venus-like exoplanets with high H$_2$O and HCl content in their atmospheres could exhibit depleted O$_3$ concentrations in their observable atmosphere (Figure \ref{fig:columnintegratedabundances}). However, intermediate H$_2$O mixing ratios in the lower-atmosphere, 10$^{-7}$ for the M dwarf case, could enhance O$_3$ concentrations in the upper atmosphere.

\textit{Atmospheric Pressure-Temperature Structure}: Venus-like exoplanets are expected to exhibit a diversity of orbital separations \citep{Williams2002}, and climate regimes \citep{Williams2003}. Again, this motivates an understanding of the implications of this diversity for the O$_3$ content of Venus-like exoplanet atmospheres. Within the paradigm of our modeling framework, Venus-like exoplanets with hotter observable upper atmospheres atmospheres will have lower O$_3$ concentrations in the same region (Figure \ref{fig:columnintegratedtemperature}).

\textit{Integrated Incident Stellar Flux}: Given the large range of orbital separations that are consistent with planets being potentially Venus-like \citep{Williams2002,Turbet2023}, it follows that Venus-like exoplanets could experience a wide variety of IISFs. Our results can therefore be used to estimate the O$_3$ content in the upper atmospheres of Venus-like exoplanets with different orbital separations. These results show that, within the paradigm of our modeling framework, Venus-like exoplanets that experience greater IISFs will have greater O$_3$ concentrations in their upper atmospheres. However, it is important to note that the increase in O$_3$ concentrations due to a greater IISF would be countered by the decrease in O$_3$ concentrations caused by the additional atmospheric heating associated with higher IISFs.

However, the most pertinent question to ask is: do any of these 1D photochemical models spanning the parameter space for Venus-like exoplanets achieve observable O$_3$ concentrations in their upper atmosphere? To estimate the minimum O$_3$ column density that would be observable with JWST, we perform a column integration of an isochemical O$_3$ profile between 70\,km and 130 \,km, using the number density profile from our fiducial Venus model. We choose 10$^{-6.5}$ as the O$_3$ mixing ratio for the isochemical O$_3$ profile, which is the mixing ratio from the model in \cite{Tremblay2020}, in which they argue that the O$_3$ would be detectable in an Earth-like atmosphere with 50 transits around an M star with JWST. This calculation yields a value of 5.5$\times$10$^{17}$ cm$^{-2}$ for the estimated minimum observable O$_3$ column density, which is show in figures \ref{fig:columnintegratedabundances}-\ref{fig:columnintegratedIISF}. This is only an order of magnitude less than the average O$_3$ column density in Earth's atmosphere: 8.8$\times$10$^{17}$ cm$^{-2}$ \citep{Staehelin2001}.

None of our 1D models achieve O$_3$ column densities within an order of magnitude of 5.5$\times$10$^{17}$ cm$^{-2}$: the threshold required for O$_3$ detectability according to \cite{Tremblay2020}. Furthermore, the threshold for the observable column density of O$_3$ would likely be higher than 5.5$\times$10$^{17}$ cm$^{-2}$ for Venus-like exoplanets orbiting K and G stars, given the difficulty in observing rocky exoplanets around high-mass stars \citep{Scalo2007}. Therefore, within our modeling framework, there is no location within the parameter space spanned by Venus-like exoplanets with observable atmospheric O$_3$ concentrations. 

The lack of observable O$_3$ concentrations within the parameter space spanned by our 1D photochemical models leads us to ask: can the unknown chemical pathway producing O$_3$ in Venus's atmosphere produce observable O$_3$ in the atmospheres of Venus-like exoplanets? For the 1D model shown in figure \ref{fig:ghosto3nophoto} with O$_3$ production rates of 10$^5$ cm$^{-3}$s$^{-1}$, the resulting O$_3$ column density between 70\,km and 130\,km is 1.25$\times$10$^{17}$ cm$^{-2}$. If the unknown chemical pathway were operating in Venus-like exoplanet atmospheres at a slightly higher rate than on Venus, it would produce observable O$_3$. Therefore, an understanding of this unknown chemical pathway is required to fully explore the question of abiotic O$_3$ production on Venus-like exoplanets.

\subsection{Unknown Chemical Pathway Operating on Venus-like Exoplanets}
\label{subsec:newchemistryexoplanets}

Our limited understanding of the origin of Venus's mesospheric O$_3$ layer limits our ability to predict O$_3$ concentrations on Venus-like exoplanets. Nevertheless, if this mechanism produces O$_3$ on the nightside of a Venus-like exoplanet at rates comparable to our model (Figure \ref{fig:ghosto3nophoto}), the resulting O$_3$ is unlikely to survive on the dayside due to photochemical destruction, as demonstrated in Figure \ref{fig:ghosto3photo}. However, given that transmission spectroscopy probes the terminator regions of a planet's atmosphere \citep{Seager2000}, it is pertinent to ask the questions: will the O$_3$ from the nightside survive long enough in the terminator regions to be detectable with transmission spectroscopy? While solar occultation experiments failed to detect O$_3$ in Venus's terminator \citep{Vandaele2008}, it remains unclear whether O$_3$ could persist in the terminator region of a Venus-like planet orbiting a K- or M-type star.

To answer this question, we simulate a 0D gas parcel circulating from the nightside to the dayside atmosphere of our model with the arbitary O$_3$ flux, to estimate the timescales over which the O$_3$ produced on the nightside is destroyed by photochemistry. We perform 0D chemical kinetics simulations, using the mixing ratios at 100\,km from our nightside 1D model with an ozone flux of 10$^5$ cm$^{-3}$ s$^{-1}$ at 64\,km as input. Photochemical reactions are included in these models, with a stellar flux density at 100\,km computed via radiative transfer using the mixing ratios from our fiducial model for a planet orbiting an M star. The M star was chosen for these simulations given the interest in terrestrial exoplanets orbiting M stars \citep{Scalo2007}. Several of these simulations were performed with increased input O$_3$ mixing ratios, ranging from 10$^{-7}$ to 10$^{-2}$, to determine if larger nightside O$_3$ production rates are required for the O$_3$ to survive at the terminator. An input O$_3$ mixing ratio of 10$^{-7}$ corresponds to an O$_3$ production rate of 10$^5$ cm$^{-3}$ s$^{-1}$.

\begin{figure}
    \includegraphics[width=0.5\textwidth]{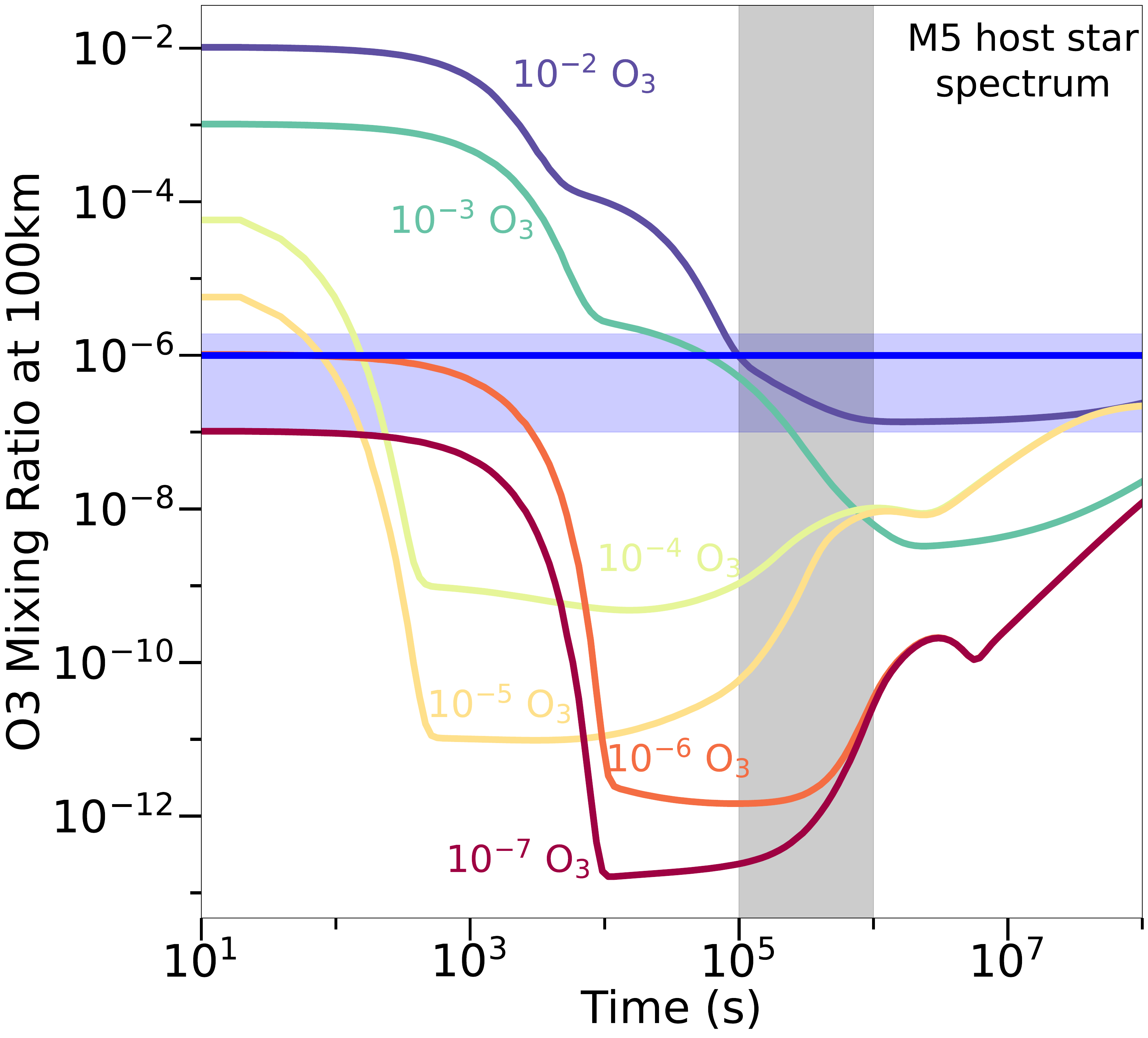}
    \caption{O$_3$ mixing ratio as a function of simulation time for 0D chemical-kinetics simulations, where the output mixing ratios at 100 km from our nightside model with an O$_3$ flux of 10$^5$ cm$^{-3}s^{-1}$ at 64\,km as input. These simulations incorporate photochemistry, with an M star stellar spectrum, to model the evolution of O$_3$ produced in the nightside atmosphere at 100 km as it circulates toward the dayside. Each line corresponds to a different O$_3$ mixing ratio used as input. The blue region corresponds to the observed O$_3$ mixing ratio at 100\,km with its reported uncertainty \citep{Montmessin2011} and the grey region corresponds to the range of atmospheric circulation timescales for Venus's atmosphere, using minimum and maximum wind speeds of 10 ms$^{-1}$ and 1000 ms$^{-1}$ respectively.}
    \label{fig:ozonedaysidecirculation}
\end{figure}

Figure \ref{fig:ozonedaysidecirculation} shows the results of these simulations. For the model with an initial O$_3$ mixing ratio of 10$^{-6}$, corresponding to the nightside photochemical model with an O$_3$ flux of 10$^5$ cm$^{-3}s^{-1}$, photochemical reactions destroy the O$_3$ produced on the nightside over a timescale of hours. The O$_3$ then reforms due to increased O production from O$_3$ photolysis. Also, if O$_3$ were being produced on the nightside at 10\,ppm or 100\,ppm, then the O$_3$ would only survive over a timescale of minutes. Therefore, if O$_3$ were being produced at ppm concentrations on the nightside of a Venus-like exoplanet orbiting an M star, it is possible that the O$_3$ could survive long enough at the terminator to be observable in transmission spectroscopy with JWST. Models of synthetic JWST transmission spectra for Venus-like exoplanets are needed to confirm this possibility.


\section{Conclusions}
\label{sec:conclusions}

In the era of JWST, ozone remains a promising disambiguator between Earth-like and Venus-like exoplanets and candidate biosignature, given its formation as a by-product of photosynthetically produced oxygen and its prominent absorption feature in the mid-infrared. However, its utility in both these cases is complicated by the presence of an ozone layer in Venus's observable above-cloud atmosphere, with ozone concentrations comparable to those found in Earth’s atmosphere. If a similar O$_3$ layer were to exist on a Venus-like exoplanet, it could be as detectable as an O$_3$ layer on an Earth-like planet, limiting its use as a biosignature and a disambiguator between Earth-like and Venus-like exoplanets. To address this issue, we have employed 1D photochemical models of Venus's atmosphere to identify the main factors influencing ozone production in the atmospheres of Venus-like planets and to shed light on the origin of Venus's mesospheric ozone layer, assessing the likelihood of this false-positive scenario.


A key result from our study is that O$_3$ formation on Venus's nightside, induced by fluxes of O atoms from the dayside, is unlikely to elevate O$_3$ concentrations to those required by observations \citep{Montmessin2011}, contrary to the findings in \cite{Krasnopolsky2013}. This result reopens the question as to the origin of this ozone. Our modelling demonstrates that none of lower-atmosphere chemistry, thermochemistry and photochemistry alone can raise the O$_3$ mixing ratio in the mesosphere to that required by observations without also violating the observed upper limit of O$_2$ reported by \cite{Mills1999}. Therefore, we propose the existence of an unknown chemical pathway operating in Venus's atmosphere that produces the bulk of the ozone in Venus's mesosphere. This pathway must satisfy three criteria to align with observational constraints: it must produce ppm levels of O$_3$, it must generate O$_3$ either in or above the cloud layer, and it must not result in O$_2$ concentrations in excess of the observed upper limit for the O$_2$ mixing ratio. Unless there is also an additional chemical sink of O$_2$ in the atmosphere, these criteria require that the O$_3$ be produced on the nightside at a rate of 10$^5$-10$^7$ cm$^{-3}$ s$^{-1}$.

The exact nature of the chemical pathway producing ozone in Venus's observable atmosphere remains an open question. While our analysis suggests that lightning is unlikely to generate sufficient O$_3$ to match observations, alternative forms of exotic chemistry may be at play in Venus's cloud layer, including droplet chemistry \citep{Rimmer2021} and even potential biological processes \citep{Greaves2021}. Future photochemical models should incorporate these potential chemical pathways and assess their ability to produce the required O$_3$. However, to fully address the question of O$_3$ in Venus's atmosphere, more direct observations of Venus's mesosphere are essential, particularly from upcoming missions such as DAVINCI \citep{Garvin2022}, that can measure O$_3$ itself and provide evidence of the chemical pathways that could theoretically produce the required O$_3$. Furthermore, direct measurements of O$_2$ mixing ratios in Venus's mesosphere by these missions would provide valuable insight into the nature of these chemical pathways, especially since O$_2$ nightglow observations \citep{Evdokimova2025} cannot directly detect O$_2$.

Understanding the mechanism responsible for O$_3$ production in Venus's atmosphere is required for accurately predicting O$_3$ concentrations in the upper atmosphere of a Venus-like exoplanet, since our results show that none of the conventional chemical processes explored in this study can achieve observable O$_3$ concentrations in the upper atmosphere on their own. Once the mechanism is identified for Venus, future research should investigate how it operates across the parameter space of Venus-like worlds. If this mechanism functions similarly in a Venus-like exoplanet, it would produce potentially observable O$_3$ concentrations primarily in the nightside atmosphere. This scenario makes an O$_3$ false-positive less likely for direct imaging missions such as HWO and LIFE, but it remains a possibility for transmission spectroscopy with JWST. 



\section*{Acknowledgments}

The authors thanks Joanna Petkowska-Hankel for the background artwork used in figure \ref{fig:ozoneschematic}. R.C. thanks the Science and Technology Facilities Council (STFC) for the PhD studentship (grant reference ST/Y509139/1). O.S. acknowledges support from STFC grant UKRI1184. S.J. acknowledges funding support from ETH Zurich and the NOMIS Foundation in the form of a research fellowship. P.B.R. acknowledges support from the Cambridge Initiative for Planetary Science and Life in the Universe (IPLU). The NOMIS Foundation ETH Fellowship Programme and respective research are made possible thanks to the support of the NOMIS Foundation. T.C. thanks the Science and Technology Facilities Council (STFC) for the PhD studentship (grant reference ST/X508299/1). We also thanks Alex Archibald for his advice relating to the atmospheric chemistry of ozone. Finally, we thank the anonymous referee for helpful and insightful comments that helped improve the quality of this manuscript.

\section*{Data Availability}

The data underlying this article will be shared on reasonable request to the corresponding author.


\bibliographystyle{mnras}
\bibliography{references.bib} 



\appendix
\label{app:mainappendix}

\begin{figure*}
    \includegraphics[width=\textwidth]{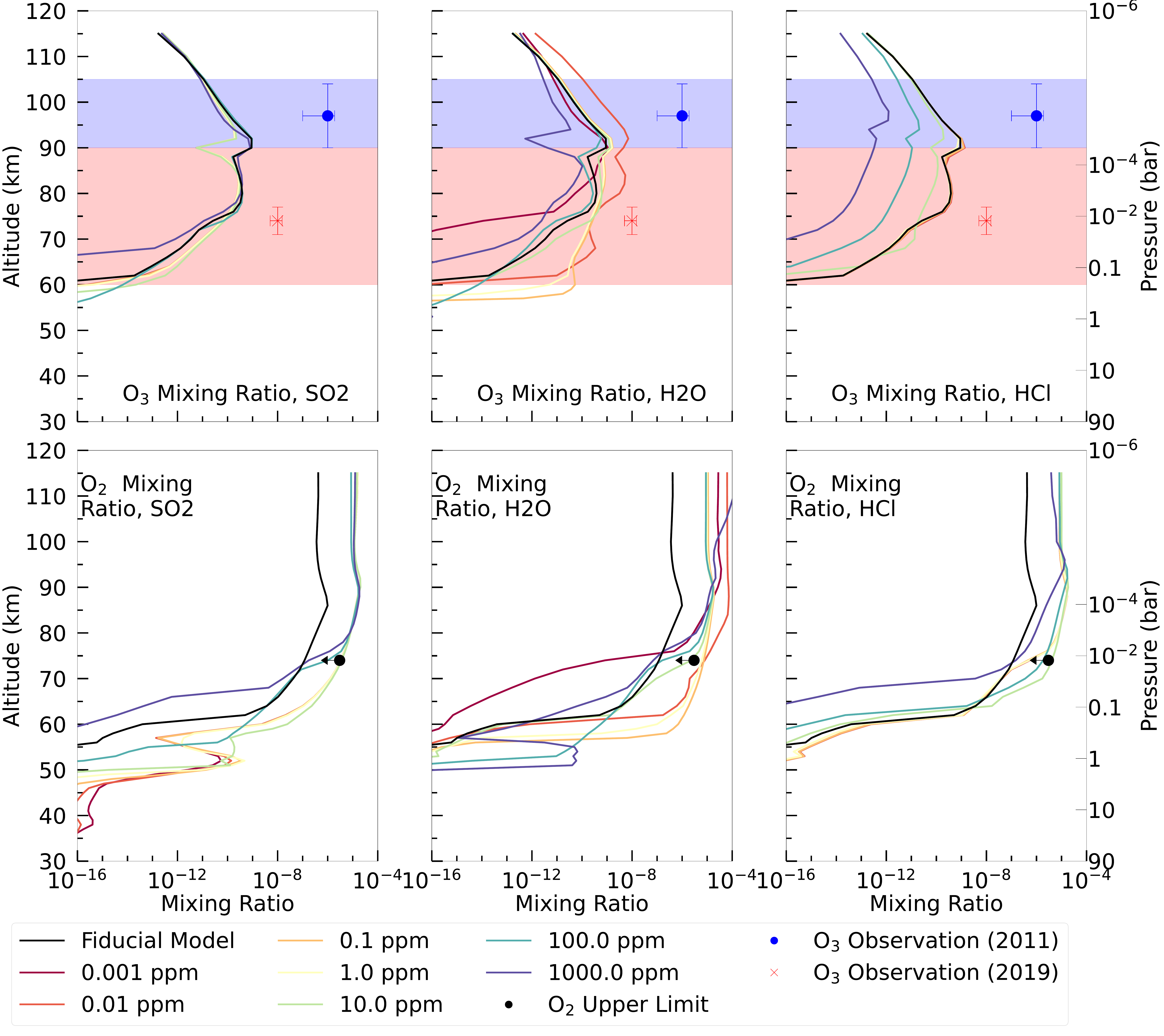}
    \caption{Ozone ($\mathrm{O}_3$) mixing ratios (top row) and molecular oxygen ($\mathrm{O}_2$) mixing ratios (bottom row) as a function of altitude for photochemical models with varying lower-atmosphere mixing ratios of H$_2$O, SO$_2$, and HCl. Each line represents a different lower-atmosphere mixing ratio for the respective species, with the black lines indicating the $\mathrm{O}_3$ and $\mathrm{O}_2$ mixing ratios for the fiducial Venus model. Each column corresponds to a specific species for which the lower-atmosphere mixing ratio is varied. Data points are shown for the observations from \protect\cite{Montmessin2011} (blue) and \protect\cite{Marcq2019} (red), with shaded regions marking the altitude ranges of these observations.}
    \label{fig:differentsurfaceabundances}
\end{figure*}

\begin{figure*}
    \includegraphics[width=\textwidth]{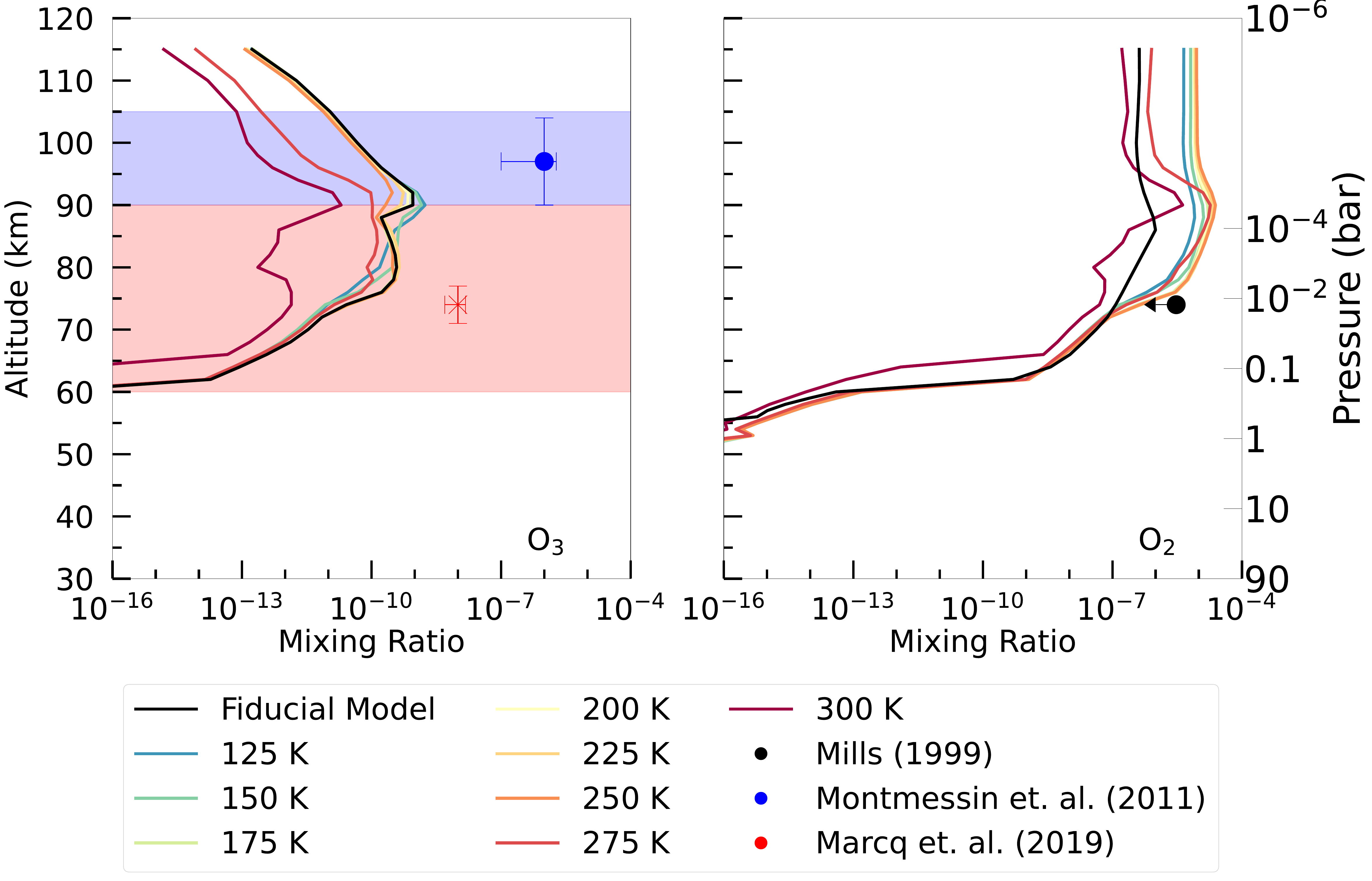}
    \caption{Ozone ($\mathrm{O}_3$) mixing ratios (left) and molecular oxygen ($\mathrm{O}_2$) mixing ratios (right) as a function of altitude for photochemical models using different pressure-temperature (PT) profiles. Each line represents a different PT profile, with varying fixed temperatures above 78 km. Data points are shown for the observations from \protect\cite{Montmessin2011} (blue) and \protect\cite{Marcq2019} (red), with shaded regions marking the altitude ranges of these observations.}
    \label{fig:O$_3$mrdifferentptprofiles}
\end{figure*}

\begin{figure*}
    \includegraphics[width=\textwidth]{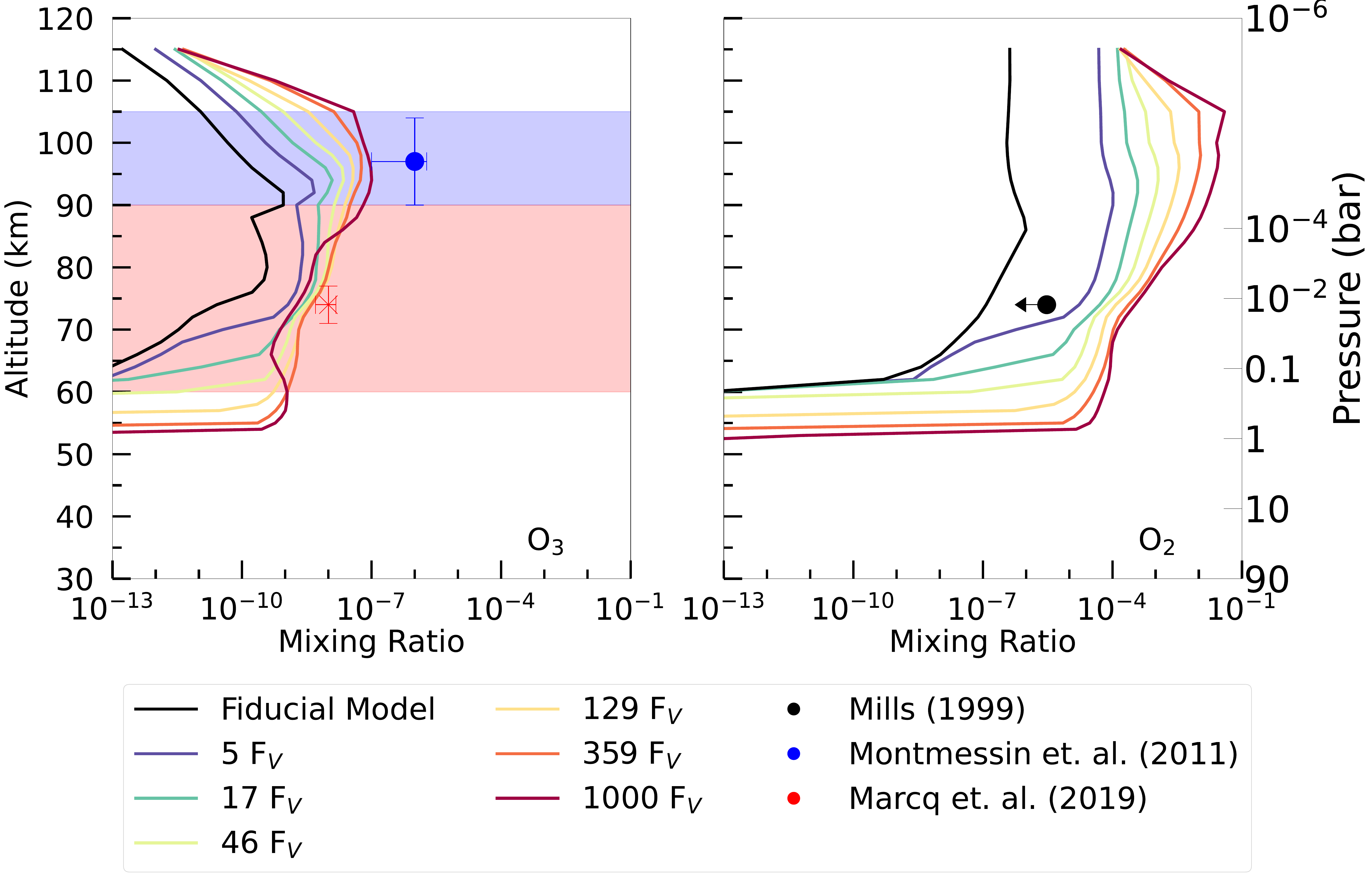}
    \caption{Ozone ($\mathrm{O}_3$) mixing ratios (left) and molecular oxygen ($\mathrm{O}_2$) mixing ratios (right) as a function of altitude for photochemical models with different integrated incident stellar fluxes (IISF). Each line corresponds to a different IISF. Data points are shown for the observations from \protect\cite{Montmessin2011} (blue) and \protect\cite{Marcq2019} (red), with shaded regions marking the altitude ranges of these observations.}
    \label{fig:O$_3$mrdifferentptprofiles}
\end{figure*}

\begin{figure}
    \includegraphics[width=0.5\textwidth]{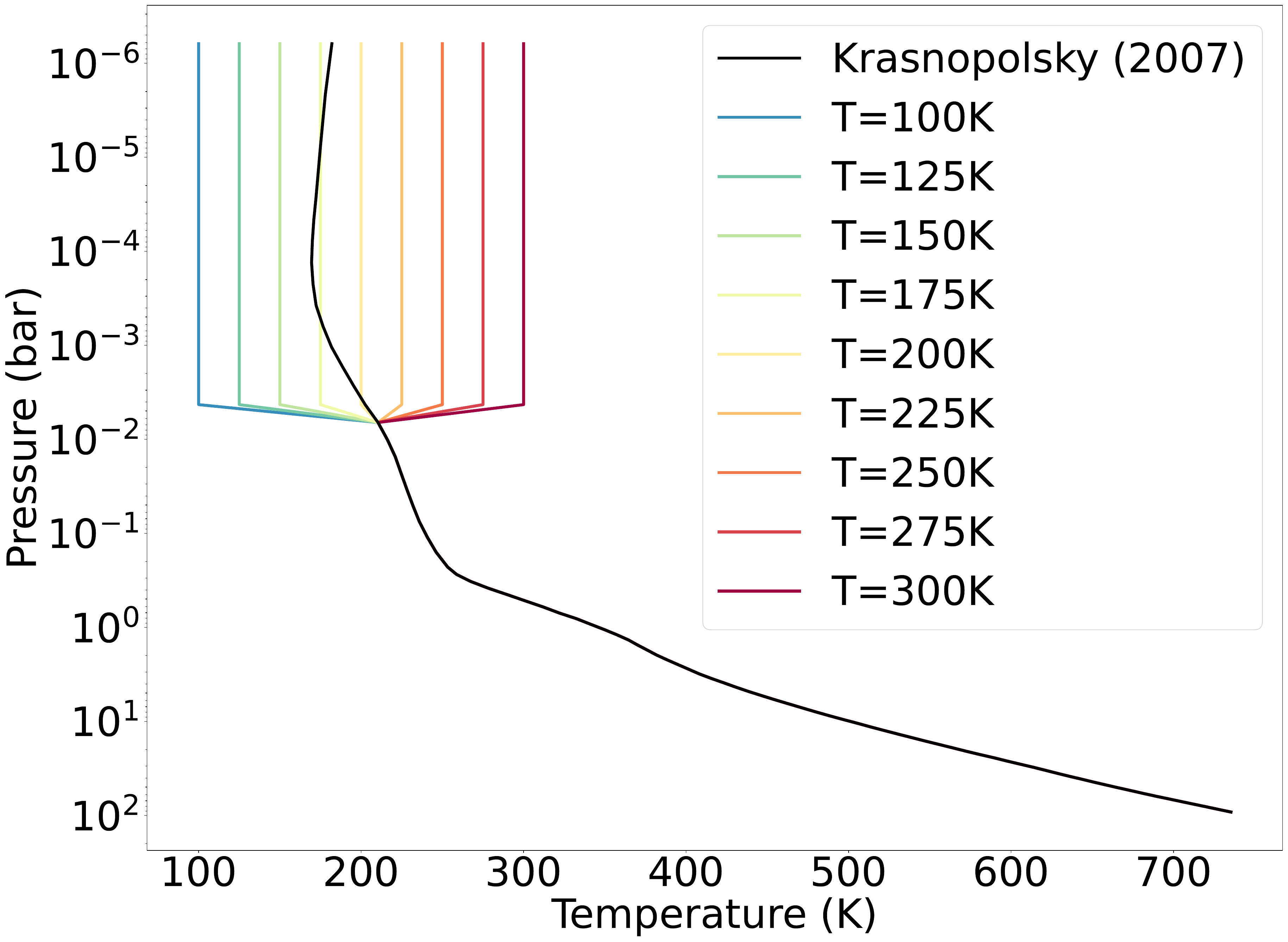}
    \caption{Pressure-temperature (PT) profiles used to investigate the effect of thermochemistry on O$_3$ production. All models share the same PT profile shown in figure \ref{fig:ptprofile} below 78\,km. Above this altitude, the profiles are isothermal, with a different isothermal temperature assigned to each model.}
    \label{fig:differentPTprofiles}
\end{figure}


\bsp	
\label{lastpage}
\end{document}